\documentclass[twoside,jaspp]{ametsoc} 

\input epsf 
\usepackage{times} 

\lefthead{Basu and Port\'{e}-Agel}   
\righthead{LES of  SBL}  
\received{February, 2005}

\sectionnumbers  
\printfigures   

\authoraddr{Sukanta Basu, St.  Anthony Falls Laboratory, University of
Minnesota, Minneapolis, MN 55414 (basus@msi.umn.edu)}

\begin{document}
\bibliographystyle{ametsoc}

\title{Large-Eddy Simulation of Stably Stratified Atmospheric Boundary
Layer Turbulence: A Scale-Dependent Dynamic Modeling Approach}

\author{Sukanta Basu and  Fernando Port\'{e}-Agel} \affil{St.  Anthony
Falls Laboratory, University of Minnesota, Minneapolis, MN 55414}

\begin{abstract}
A  new  tuning-free   subgrid-scale  model,  termed  `locally-averaged
scale-dependent dynamic'  (LASDD) model, is  developed and implemented
in  large-eddy simulations (LESs) of  stable boundary  layers. The  new 
model dynamically  computes  the  Smagorinsky  coefficient and  the
subgrid-scale  Prandtl  number based  on  the  local  dynamics of  the
resolved  velocity  and temperature  fields.   Overall, the  agreement
between  the  statistics  of  the LES-generated  turbulence  and  some
well-established  empirical formulations  and  theoretical predictions
(e.g.,  Nieuwstadt's  local scaling  hypothesis)  is  remarkable.  The
results  show   clear  improvements  over  most   of  the  traditional
subgrid-scale models  in the surface  layer. Moreover, in  contrast to
previous large-eddy  simulations of  stable boundary layers  that have
strong  dependence  on   grid  resolution,  the  simulated  statistics
obtained  with  the  LASDD  model show  relatively  little  resolution
dependence for the range of grid sizes considered here. In essence, we
show   that  the   new   LASDD  model   is   a  robust   subgrid-scale
parameterization  for  reliable,  tuning-free  simulations  of  stable
boundary layers, even with relatively coarse resolutions.
\end{abstract}

\begin{article}

\section{Introduction}
Atmospheric boundary  layers (ABLs) are usually  classified into three
types: neutral, convective and  stable, based on atmospheric stability
(buoyancy effects) and the dominant mechanism of turbulence generation
(\citeauthor{stul88}       \citeyear{stul88};      \citeauthor{arya01}
\citeyear{arya01}).   The  boundary  layer becomes  stably  stratified
whenever the  underlying surface is  colder than the air.   Under this
atmospheric condition, turbulence is  generated by shear and destroyed
by  negative  buoyancy and  viscosity.   Because  of this  competition
between shear and buoyancy effects,  the strength of turbulence in the
stable  boundary layer  (SBL)  is  much weaker  in  comparison to  the
neutral  and convective  boundary  layers.  As  a  result, the  stable
boundary  layer is also  much shallower  and characterized  by smaller
eddy motions.  Stable boundary  layer turbulence has not received much
attention despite  its scientifically intriguing  nature and practical
significance (e.g., numerical weather prediction -- NWP, and pollutant
transport).  This might be attributed to the lack of adequate field or
laboratory measurements,  to the inevitable  difficulties in numerical
simulations   (arising   from   small   scales  of   motion   due   to
stratification) and to the intrinsic complexity in its dynamics (e.g.,
occurrences  of intermittency,  Kelvin-Helmholtz  instability, gravity
waves,      low-level     jets,      meandering      motions     etc.)
(\citeauthor{hunt96}~\citeyear{hunt96};
\citeauthor{mahr98a}~\citeyear{mahr98a};
\citeauthor{derb99}~\citeyear{derb99}).    Not   surprisingly,  today,
there is a general  consensus among researchers that our understanding
of the  stable boundary layer  (especially the very stable  regime) is
quite          poor          (\citeauthor{mahr98a}~\citeyear{mahr98a};
\citeauthor{derb99}~\citeyear{derb99};
\citeauthor{holt03}~\citeyear{holt03}) and `even small future advances
justify more work' \cite{mahr98a}.

In order to improve our understanding of SBL turbulence and to explore
some of its  inherent characteristics, in this study we  make use of a
contemporary   numerical  modeling   approach,  known   as  large-eddy
simulation.   Following   the  pioneering  works  of  
Deardorff~\cite{dear70b,dear72,dear74,dear80} 
over  the years LES  has become an
indispensable      tool     to      study      the     ABL      (e.g.,
\citeauthor{moen84}~\citeyear{moen84};
\citeauthor{nieu91}~\citeyear{nieu91};
\citeauthor{andr94}~\citeyear{andr94};
\citeauthor{maso94}~\citeyear{maso94};
\citeauthor{sull94}~\citeyear{sull94};
\citeauthor{koso97}~\citeyear{koso97};
\citeauthor{albe99}~\citeyear{albe99};
\citeauthor{port00}~\citeyear{port00};
\citeauthor{bear04b}~\citeyear{bear04b}).   However,  until   now  LES
models  have  not  been   sufficiently  faithful  in  reproducing  the
characteristics of moderately and strongly stable atmospheric boundary
layers (\citeauthor{saik00} \citeyear{saik00}; \citeauthor{holt03} 
\citeyear{holt03}).
The main weakness of  LES is associated
with our limited  ability to accurately account for  the dynamics that
are  not  explicitly  resolved   in  the  simulations.   Under  stable
conditions -- due to flow stratification -- the characteristic size of
the eddies  becomes increasingly smaller with  increasing atmospheric
stability, which  eventually imposes an  additional burden on  the LES
subgrid-scale (SGS) models.  The recent GABLS (Global Energy and Water
Cycle Experiment Atmospheric Boundary Layer Study) LES intercomparison
study \cite{bear04b} highlights that LESs of moderately stable BLs are
quite sensitive to SGS models  at a relatively fine resolution of 6.25
m.  At  a coarser resolution (12.5  m), a couple of  commonly used SGS
models even laminarized  spuriously.  Occasionally, laminarization was
manifested  by  a  near-linear  (without  any  curvature)  temperature
profile; at times the SGS  contributions to the total momentum or heat
fluxes were larger than fifty  percent in the interior of the boundary
layer.  This breakdown of traditional SGS models undoubtedly calls for
improved SGS  parameterizations in order  to make LES a  more reliable
tool to  study stable  boundary layers. The  present study  is devoted
towards this goal.

The organization of this paper is as follows. In Section 2, we briefly
describe  the basic  philosophy  of large-eddy  simulation. The  newly
developed   locally-averaged  scale-dependent   dynamic   (LASDD)  SGS
modeling  approach is presented  in Section  3. Simulations  of stably
stratified  atmospheric  boundary  layers  are  presented  in  Section
4. Lastly, in  Section 5, we  summarize our research and  elaborate on
the prospects for future research in this subject area.

\section{Subgrid-Scale Modeling and SGS Parameter Estimation}

In rotation-influenced ABLs,  the equations governing the conservation
of momentum (using the Boussinesq approximation) and temperature are:
\begin{equation}\label{NSa}
\frac{\partial     \tilde{u}_i}{\partial    t}     +    \frac{\partial
(\tilde{u}_i\tilde{u}_j)}{\partial      x_j}     =     -\frac{\partial
\tilde{p}}{\partial  x_i}   -  \frac{\partial\tau_{ij}}{\partial  x_i}
+\delta_{i3}g\frac{(\tilde{\theta}  -  \langle \tilde{\theta}  \rangle
)}{\theta_0} + f_c\epsilon_{ij3}\tilde{u}_j + F_i
\end{equation}
\begin{equation}\label{NSb}
\frac{\partial\tilde{\theta}}{\partial               t}              +
\frac{\partial(\tilde{u}_j\tilde{\theta})}{\partial     x_j}    =    -
\frac{\partial q_j}{\partial x_j},
\end{equation}
where  $t$   is  time,  $x_j$   is  the  spatial  coordinate   in  the
$j$-direction,  $u_j$ is  the  velocity component  in that  direction,
$\theta$  is potential  temperature, $\theta_0$  is  reference surface
potential temperature,  $p$ is dynamic pressure,  $\delta_{i3}$ is the
Kronecker  delta, $\epsilon_{ij3}$ is  the alternating  unit tensor,
$g$ is the gravitational acceleration, $f_c$ is the Coriolis parameter
and $F_i$  is a forcing term  (e.g., Geostrophic wind  or imposed mean
pressure  gradient).  Molecular  dissipation and  diffusion  have been
neglected since  the Reynolds number  of the ABL  is very high  and no
near-ground viscous processes  are resolved.  The $\langle~\rangle$ is
used to define a  horizontal plane average. The $\widetilde{(\cdots)}$
denotes   a   spatial  filtering   operation,   using   a  filter   of
characteristic  width  $\Delta_f$. These  filtered  equations are  now
amenable  for  numerical  solution  (LES)  on  a  grid  of  mesh  size
$\Delta_g$, considerably  larger than the smallest  scale of turbulent
motion (the so-called Kolmogorov scale).  It is common practice to use
the ratio of filter-width to grid-spacing, $\Delta_f/\Delta_g =  1$ or 2
[see  the  Chapter 9  of  \citeauthor{geur03} (\citeyear{geur03})  for
detailed   discussion  on  this   ratio  and   its  impact   on  error
dynamics]. In this study, we use a ratio of 2.

The effects of the unresolved  scales (smaller than $\Delta_f$) on the
evolution  of $\tilde{u}_i$  and  $\tilde{\theta}$ appear  in the  SGS
stress  $\tau_{ij}$ (see Equation  \ref{NSa}) and  the SGS  flux $q_i$
(see Equation \ref{NSb}), respectively. They are defined as
\begin{equation}\label{SGSa}
\tau_{ij} = \widetilde{u_i u_j}-\tilde{u}_i \tilde{u}_j
\end{equation}
and
\begin{equation}\label{SGSb}
q_i = \widetilde{u_i\theta} - \tilde{u}_i\tilde{\theta}.
\end{equation}
Note that the  SGS stress and flux quantities are  unknown and must be
parameterized  (using a  SGS  model)  as a  function  of the  resolved
velocity and temperature fields.

Eddy-viscosity models, the most  popular SGS models, use the `gradient
hypothesis' and formulate the  $ij$-component of the SGS stress tensor
(the deviatoric part) as follows (\citeauthor{smag63} \citeyear{smag63};
\citeauthor{geur03} \citeyear{geur03}):
\begin{equation}\label{EddyVisc}
\tau_{ij} -\frac{1}{3}\tau_{kk}\delta_{ij} = -2\nu_t\tilde{S}_{ij},
\end{equation}
where
\begin{eqnarray*}
\tilde  {S}_{ij}  =  \frac{1}{2}\left(  {\frac{\partial  \tilde  {u}_i
}{\partial  x_j  }  +   \frac{\partial  \tilde  {u}_j  }{\partial  x_i
}}\right)\nonumber
\end{eqnarray*}
is  the   resolved  strain  rate   tensor  and  $\nu_t$   denotes  the
eddy-viscosity.  It  is well-known that the  eddy-viscosity SGS models
give  a poor  prediction of  the SGS  stresses on  a local  level (see
\citeauthor{sarg99}       \citeyear{sarg99};       \citeauthor{mene00}
\citeyear{mene00};   \citeauthor{geur03}   \citeyear{geur03}.)   Their
underlying  assumption of  strain  rates being  aligned  with the  SGS
stress     tensor    is    unrealistic     (see    \citeauthor{higg03}
\citeyear{higg03} and the references therein).  Furthermore, these SGS
models are purely  dissipative, i.e., they do not  allow local reverse
energy transfer  (known as `backscatter')  \cite{sarg99}.  Despite all
these deficiencies,  without any doubt, the  eddy-viscosity models are
the most  commonly used SGS  models in the atmospheric  boundary layer
community.

From  dimensional   analysis, the  eddy-viscosity   ($\nu_t$)  can  be
interpreted as  the product of  a characteristic velocity scale  and a
characteristic  length scale \cite{geur03}.   Different eddy-viscosity
formulations basically  use different velocity and  length scales. The
most  popular  eddy-viscosity  formulation  is the  Smagorinsky  model
\cite{smag63}:
\begin{equation}
\nu_t = (C_S\Delta_f)^2 \left|\tilde{S}\right|,
\end{equation}
where  $C_S$  is  the  so-called  Smagorinsky  coefficient,  which  is
adjusted   empirically   or   dynamically   to  account   for   shear,
stratification and grid-resolution, and
\begin{eqnarray*}
| \tilde    {S}|   =    \left({2\tilde   {S}_{ij}    \tilde   {S}_{ij}
}\right)^{1/2}\nonumber
\end{eqnarray*}
is the magnitude  of the resolved strain rate  tensor.  In contrast to
the  Smagorinsky-type  eddy-viscosity  model, the  turbulence  kinetic
energy (TKE) based  eddy-viscosity model utilizes (\citeauthor{moen84}
\citeyear{moen84};        \citeauthor{sull94}       \citeyear{sull94};
\citeauthor{sull03} \citeyear{sull03}):
\begin{equation}
\nu_t = C_K l E_{SGS}^{1/2},
\end{equation}
where  $C_K$ is  a modeling  coefficient, $l$  is a  length  scale and
$E_{SGS}$ is the SGS turbulence kinetic energy. This modeling approach
involves solving an  extra prognostic equation for the  SGS TKE. Based
on     the    Kolmogorov's    scaling     laws,    \citeauthor{wong94}
(\citeyear{wong94}) proposed yet another eddy-viscosity model:
\begin{equation}
\nu_t = C^{2/3}\Delta_f^{4/3}\epsilon^{1/3} = C_\epsilon\Delta_f^{4/3},
\end{equation}
where $\epsilon$ is the dissipation rate of energy and $C_\epsilon$ is
a  model  coefficient to  be  specified  (or determined  dynamically).
There are  numerous other formulations for  eddy-viscosity existing in
the    literature   (e.g.,   the    Structure   Function    model   of
\citeauthor{meta92}~\citeyear{meta92}).  An  extensive review of these
formulations is given by \citeauthor{saga01} (\citeyear{saga01}).

Similar to the SGS stresses, the  SGS heat fluxes are modeled with the
eddy-diffusivity models as:
\begin{equation}
q_i=-\nu_{ht}\frac{\partial     \tilde{\theta}}{\partial     x_i}    =
-\frac{\nu_t}{Pr_{SGS}}\frac{\partial \tilde{\theta}}{\partial x_i},
\end{equation}
where $Pr_{SGS}$ is the SGS Prandtl number.

The  values of  the Smagorinsky-type  SGS model  parameters  $C_S$ and
$Pr_{SGS}$ are well  established for homogeneous, isotropic turbulence
\cite{lill67}.  However, the value  of $C_S$ is expected to decrease 
with increasing mean  shear and stratification. This has been confirmed 
by various recent field studies (\citeauthor{port01} \citeyear{port01};
\citeauthor{klei03}       \citeyear{klei03};       \citeauthor{sull03}
\citeyear{sull03}; \citeauthor{klei04a} \citeyear{klei04a}).  In order
to  account for  this, application  of the  traditional eddy-viscosity
model  in LES  of ABL  flows (with  strong shear  near the  ground and
temperature-driven stratification) has  traditionally involved the use
of various types of  wall-damping functions and stability corrections,
which are either based on the phenomenological theory of turbulence or
empirically  derived  from   observational  data  \cite{maso94}.   For
example,  recently, based  on  the HATS  (Horizontal Array  Turbulence
Study)  field  campaign  data \citeauthor{klei03}  (\citeyear{klei03},
\citeyear{klei04a}) proposed the following empirical form for $C_S$:
\begin{equation}
(C_S)_{\Delta_f}                                                      =
c_0\left[1+R\left(\frac{\Delta_f}{L}\right)\right]^{-1}
\left[1+\left(\frac{c_0}{\kappa}\frac{\Delta_f}{z}\right)^n\right]^{1/n},
\end{equation}
where $L$ is the Obukhov  length, $\kappa$ is the von Karman constant,
$R$ is  the ramp function, $n$  = 3 and $c_0  \approx 0.135$.  Another
example  would be  Smagorinsky-type SGS  models that  impose  both the
wall-damping  and  stability corrections  based  on  the Kansas  field
experiment    data    (see   \citeauthor{brow94}    \citeyear{brow94};
\citeauthor{maso94}       \citeyear{maso94};       \citeauthor{maso99}
\citeyear{maso99};  \citeauthor{bear04a} \citeyear{bear04a}).   In the
case  of TKE-based  eddy-viscosity  models, the  length  scale $l$  is
usually  set equal  to the  filter width  $\Delta_f$ for  unstable and
neutral  stratifications  and equal  to  $C_l  \sqrt{E_{SGS}} /N$  for
stable    stratification    (\citeauthor{sull94}    \citeyear{sull94};
\citeauthor{saik00}       \citeyear{saik00};       \citeauthor{sull03}
\citeyear{sull03}),    following   the    suggestion    of 
\citeauthor{dear80} (\citeyear{dear80}).  
Here, $C_l$ is a  coefficient to be prescribed and $N$
is the Brunt-V\"{a}is\"{a}l\"{a} frequency.  When using this approach,
the SGS  model coefficients are  often `tuned' for different  ABL flow
conditions (\citeauthor{sull94} \citeyear{sull94}; \citeauthor{saik00}
\citeyear{saik00}; \citeauthor{sull03} \citeyear{sull03}).  There also
have  been  a few  elegant  attempts  to  derive shear  and  stability
dependent length-scales  directly from the  phenomenological theory of
turbulence (\citeauthor{hunt88} \citeyear{hunt88}; \citeauthor{schu91}
\citeyear{schu91};   \citeauthor{canu97}   \citeyear{canu97}).    `The
adequacy  of  all  these  parameterizations  for  SGS  fluxes  remains
relatively untested however' \cite{sull03}.

In the case of eddy-diffusivity SGS models, one needs to prescribe the
stability  dependence  of the  SGS  Prandtl  number ($Pr_{SGS}$).   In
general, $Pr_{SGS}$ is found  to increase under stable stratification,
which is reflected in  different SGS modeling approaches. For example,
in a  widely used  Smagorinsky-type SGS model,  the Prandtl  number is
increased from  0.44 in  the free convection  limit to 0.7  in neutral
condition  to  1.0  in  the  very  stable  regime  \cite{maso99}.   In
contrast,        the       TKE-based       SGS        model       uses
(\citeauthor{moen84} \citeyear{moen84}; \citeauthor{sull94} 
\citeyear{sull94}; \citeauthor{saik00} \citeyear{saik00}; 
\citeauthor{sull03} \citeyear{sull03}): $Pr_{SGS} =
\Delta_f/(\Delta_f+2l)$, where $l$ is defined as before.  This implies
that the  $Pr_{SGS}$ is 0.33  under convective and  neutral conditions
and  varies from  0.33 (weakly  stable) to  1.0 (very  stable)  in the
stably stratified regime.

In   summary,   most    of   the   conventional   eddy-viscosity   and
eddy-diffusivity SGS  modeling approaches involve  parameter tuning or
{\it  a priori}  prescription in  one way  or another. An  alternative
approach  is to  use the  `dynamic' SGS  modeling approach  of Germano
(\citeauthor{germ91}       \citeyear{germ91};      \citeauthor{germ92}
\citeyear{germ92};  \citeauthor{lill92}  \citeyear{lill92}).  In  this
approach, one computes the value of the unknown SGS coefficient (e.g.,
the coefficient  $C_S$ in the  Smagorinsky-type eddy-viscosity models)
dynamically at every time and position  in the flow. By looking at the
dynamics of  the flow  at two different  resolved scales  and assuming
scale similarity as well as scale invariance of the model coefficient,
one can optimize  its value.  Thus, the dynamic  model avoids the need
for {\it a priori} specification and tuning of the coefficient because
it  is evaluated  directly from  the resolved  scales in  an  LES.  In
\citeauthor{moin91}  (\citeyear{moin91}), a similar  dynamic procedure
was applied to estimate the SGS scalar flux in compressible flows.  In
essence this  procedure not  only eliminates the  need for  any ad-hoc
assumption  about  the  SGS   Prandtl  number  ($Pr_{SGS}$)  but  also
completely  decouples   the  SGS  flux  estimation   from  SGS  stress
computation, which is highly desirable.

\section{Locally-Averaged Scale-Dependent Dynamic Modeling Approach}
In  the previous  section, we  mentioned  that the  SGS stress  tensor
($\tau_{ij}$)  at  the  filter   scale  ($\Delta_f$)  is  defined  as:
$\tau_{ij} =  \widetilde{u_i u_j} -  \widetilde{u_i} \widetilde{u_j}$.
In  a seminal  work, Germano  (\citeauthor{germ91}  \citeyear{germ91};
\citeauthor{germ92}   \citeyear{germ92})   proposed   to   invoke   an
additional explicit test filter of width $\alpha \Delta_f$ in order to
dynamically  compute the SGS  coefficients.  Consecutive  filtering at
scales $\Delta_f$  and at $\alpha  \Delta_f$ leads to a  SGS turbulent
stress tensor ($T_{ij}$) at the test filter scale $\alpha \Delta_f$:
\begin{equation}
T_{ij}  = \overline{\widetilde{u_i~u_j}}  - \overline{\widetilde{u}_i}
~\overline{\widetilde{u}_j},
\end{equation}
where an  overline $\overline{(\cdots)}$ denotes filtering  at a scale
of  $\alpha  \Delta_f$.   From  the  definitions  of  $\tau_{ij}$  and
$T_{ij}$ an algebraic relation can  be formed, known in the literature
as the Germano identity:
\begin{equation}
L_{ij}       =       \overline{\widetilde{u}_i\widetilde{u}_j}       -
\overline{\widetilde{u}_i}~\overline{\widetilde{u}_j}   =   T_{ij}   -
\overline{\tau_{ij}}.
\end{equation}
This identity  is then effectively used to  dynamically obtain unknown
SGS model  coefficients.  In the  case of the Smagorinsky  model, this
identity  yields\footnote{Please  note  that  here  the  variation  of
$C_S^2$  over the  test  filter scale  has  been implicitly  neglected
\cite{germ91,lill92,vrem94}.}:
\begin{equation}
L_{ij}            -           \frac{1}{3}L_{kk}\delta_{ij}           =
\left(C_S^2\right)_{\Delta_f}M_{ij},
\end{equation}
where  $M_{ij} = 2\Delta_f^2\left(\overline{\left|\widetilde{S}\right|
\widetilde{S_{ij}}}                     -                     \alpha^2
\frac{\left(C_S^2\right)_{\alpha\Delta_f}}
{\left(C_S^2\right)_{\Delta_f}}\left|\overline{\widetilde{S}}\right|
\overline{\widetilde{S_{ij}}}\right)$.     If   one    assumes   scale
invariance,    i.e.,     $\left(C_S^2\right)_{\alpha    \Delta_f}    =
\left(C_S^2\right)_{\Delta_f}$   \cite{germ91},   then   the   unknown
coefficient  $\left(C_S^2\right)_{\Delta_f}$ can be  easily determined
following  the  error  minimization  approach  of  \citeauthor{lill92}
(\citeyear{lill92}):
\begin{equation}
\left(C_S^2\right)_{\Delta_f}   =  \frac{\langle  L_{ij}M_{ij}\rangle}
{\langle M_{ij}M_{ij}\rangle}.
\label{EqLijMij}
\end{equation}
Here  the  angular   brackets  $\langle\cdots\rangle$  denote  spatial
averaging  (\citeauthor{germ91} \citeyear{germ91}; \citeauthor{lill92}
\citeyear{lill92};        \citeauthor{zang93}       \citeyear{zang93};
\citeauthor{ghos95}       \citeyear{ghos95};       \citeauthor{port00}
\citeyear{port00}) or averaging over fluid pathlines \cite{mene96}.

In a recent  study, \citeauthor{port00} (\citeyear{port00}) found that
in a simulation of  neutral boundary layer the scale-invariant dynamic
model  is  not  dissipative  enough  in the  near-ground  region.   It
underestimates shear  in that region and also  yields excessively flat
spectra \cite{port00}. Moreover,  the dynamically computed coefficient
$C_S^2$   show  clear   scale-dependence  near   the   surface,  i.e.,
$\left(C_S^2\right)_{\Delta_f}     \neq     \left(C_S^2\right)_{\alpha
\Delta_f}$.   Similar inferences  are  also obtained  in  the case  of
passive     scalars    \cite{port04}.     Field     observations    by
\citeauthor{klei04a} (\citeyear{klei04a}) also support these results.

This prompted  \citeauthor{port00} (\citeyear{port00}) to  propose the
scale-dependent dynamic SGS model.  In this case, the scale-dependence
parameter    $\beta    =    \frac{\left(C_S^2\right)_{\alpha\Delta_f}}
{\left(C_S^2\right)_{\Delta_f}}$ is  not assumed  to be equal  to one,
rather it is determined  dynamically. In order to  implement this procedure,
one needs  to employ a second  test filtering operation at  a scale of
$\alpha^2 \Delta_f$  [denoted by $\widehat{(\cdots)}$].   Invoking the
Germano identity for the second time leads to:
\begin{equation}
Q_{ij}            -           \frac{1}{3}Q_{kk}\delta_{ij}           =
\left(C_S^2\right)_{\Delta_f}N_{ij},
\label{EqQij}
\end{equation}
where
\begin{eqnarray*}
Q_{ij}        =       \widehat{\widetilde{u_i}\widetilde{u_j}}       -
\widehat{\widetilde{u_i}}\widehat{\widetilde{u_j}} \nonumber
\end{eqnarray*}
and
\begin{eqnarray*}
N_{ij}      =     2\Delta_f^2\left(\widehat{\left|\widetilde{S}\right|
\widetilde{S_{ij}}}                     -                     \alpha^4
\frac{\left(C_S^2\right)_{\alpha^2\Delta_f}}
{\left(C_S^2\right)_{\Delta_f}}\left|\widehat{\widetilde{S}}\right|
\widehat{\widetilde{S_{ij}}}\right). \nonumber
\end{eqnarray*} 
This results in:
\begin{equation}
\left(C_S^2\right)_{\Delta_f}   =  \frac{\langle  Q_{ij}N_{ij}\rangle}
{\langle N_{ij}N_{ij}\rangle}.
\label{EqQijNij}
\end{equation} 
In  the  scale-dependent  dynamic  modeling  approach,  the  following
assumption is made:
\begin{equation}
\beta           =           \frac{\left(C_S^2\right)_{\alpha\Delta_f}}
{\left(C_S^2\right)_{\Delta_f}}                                       =
\frac{\left(C_S^2\right)_{\alpha^2\Delta_f}}
{\left(C_S^2\right)_{\alpha\Delta_f}},
\end{equation}
which is  a much weaker assumption than  the scale-invariance modeling
assumption  of $\beta =  1$.  Now,  from Equations  \ref{EqLijMij} and
\ref{EqQijNij}, one solves for the unknown parameter $\beta$, which in
turn  is  used  to compute  $\left(C_S^2\right)_{\Delta_f}$  utilizing
Equation  \ref{EqLijMij}.   Further technical  details  on this  model
could  be   found  in  \citeauthor{port00}   (\citeyear{port00}).   In
\citeauthor{port04}  (\citeyear{port04}),   similar  formulations  were
derived  for  scalars.  In  this  case, the  lumped  SGS  coefficient
$C_S^2Pr_{SGS}^{-1}$    was    determined    dynamically    and    the
scale-dependent  parameter  was  termed  as  $\beta_\theta$.   In  the
simulations of neutral boundary  layers, the scale-dependent SGS model
was found  to exhibit `right'  dissipation behavior and  more accurate
spectra in the case of  momentum \cite{port00}, as well as for passive
scalars \cite{port04}.

In  the  present study  we  found  that  the original  formulation  of
\citeauthor{port00}  (\citeyear{port00})  which involves  (horizontal)
planar   averaging  in  Equation   \ref{EqLijMij},  suffers   from  an
insufficient  SGS   dissipation  problem   in  the  outer   layers  in
simulations of  stable boundary layers.   This could be  attributed to
decoupling    between   horizontal   planes    under   stratification.
Intermittent, patchy  turbulence in  the strongly stable  outer layers
might  be another  cause. This  issue was  resolved by  using  a local
formulation  of the  scale-dependent modeling  approach, named  as the
locally-averaged  scale-dependent dynamic  (LASDD)  model.  The  model
coefficients   ($C_S^2$  and   $C_S^2  Pr_{SGS}^{-1}$)   are  obtained
dynamically  by  averaging locally  on  the  horizontal  plane with  a
stencil   of  three   by  three   grid   points.   \citeauthor{zang93}
(\citeyear{zang93}) followed a similar approach in  the scale-invariant
dynamic (i.e., $\beta = 1$) modeling of turbulent recirculating flows.
To avoid  numerical instabilities the coefficients  $C_S^2$ and $C_S^2
Pr_{SGS}^{-1}$ are  set to zero whenever the  dynamic procedure yields
negative values.  This commonly  used procedure is known as `clipping'
\cite{geur03}.    The   scale-dependence   parameters   ($\beta$   and
$\beta_\theta$) are  determined dynamically over  horizontal planes to
avoid the computational burden of computing them at every grid point 
in the flow. Solving for  $\beta$  or $\beta_\theta$ involves  a 
fifth-order  polynomial. Instead of the Newton-Raphson method used by 
\citeauthor{port00} (\citeyear{port00}) and \citeauthor{port04} 
(\citeyear{port04}), we use a more robust eigenvalue based method 
\cite{pres92} to obtain the roots of this polynomial.
In the infrequent
events that an appropriate  real root in the range of 0  to 1.2 is not
found, we  chose to invoke the scale-invariance  assumption of $\beta$
(or, $\beta_\theta$) = 1.

Please  note that, mathematically  more rigorous  (and computationally
more  expensive) local  models  are also  proposed  in the  literature
(\citeauthor{ghos95}       \citeyear{ghos95};      \citeauthor{piom95}
\citeyear{piom95};   \citeauthor{mene96}   \citeyear{mene96}).   Their
capabilities  in  the  stably  stratified atmospheric  boundary  layer
simulations have yet to be tested.

\section{LES of Stably Stratified Boundary Layers}

The  first   LES  of  a   stable  boundary  layer  was   performed  by
\citeauthor{maso90}  (\citeyear{maso90}).  They  used  the traditional
Smagorinsky-type  SGS  model  with   a  constant  SGS  Prandtl  number
($Pr_{SGS} = 0.5$). Their  results broadly supported the local scaling
hypothesis  of  Nieuwstadt  (\citeauthor{nieu84a}  \citeyear{nieu84a};
\citeauthor{nieu84b}      \citeyear{nieu84b};      \citeauthor{nieu85}
\citeyear{nieu85};  \citeauthor{derb90}  \citeyear{derb90}).  However,
one  of their  simulations showed  the run-away  cooling  problem. The
simulated surface  temperature fell  more than 30  K over  90 minutes.
\citeauthor{brow94}  (\citeyear{brow94})  repeated  these  simulations
with   their  stochastic   backscatter  SGS   model   (with  stability
corrections).   Their  results  definitely  showed  some  improvements
(especially in  the surface layer  properties) when compared  with the
Mason and Derbyshire's simulations.

\citeauthor{andr95}   (\citeyear{andr95})   simulated  weakly   stable
boundary layers  with the  TKE-based SGS model  of \citeauthor{moen84}
(\citeyear{moen84})  and also with  the two-part  eddy-viscosity model
developed  by  \citeauthor{sull94}  (\citeyear{sull94}). The  two-part
eddy-viscosity model, which is a modified version of the TKE-based SGS
model,  was  in better  agreement  with  the surface-layer  similarity
theory.

Recently,   \citeauthor{saik00}   (\citeyear{saik00})   attempted   to
simulate  a moderately stable  boundary layer  with Sullivan  et al.'s
two-part eddy-viscosity model \cite{sull94}. Although, this particular
SGS scheme  has been previously  used by Andr\'{e}n for  weakly stable
BLs \cite{andr95}, in the case of moderately stable boundary layers it
led  to  development  of  unphysical  profiles  of  various  turbulent
quantities \cite{saik00}.  The failure was  due to the collapse of SGS
vertical heat flux near the  surface.  This prompted them to propose a
two-part SGS model  to represent the SGS heat flux  similar to the SGS
momentum   flux  model  of   \citeauthor{sull94}  (\citeyear{sull94}).
However, even  after the modifications, the simulations  were found to
be too sensitive to rapid cooling \cite{saik00}.

\citeauthor{koso00}   (\citeyear{koso00})   simulated   a   clear-air,
moderately stable  boundary layer as it  approaches quasi-steady state
using    the    nonlinear    SGS    model    of    \citeauthor{koso97}
(\citeyear{koso97}).  Initial conditions  consistent with the Beaufort
Sea  Arctic Stratus  Experiment  (BASE) observations  were used.   The
first  intercomparison study  of  LES-SGS models  for stable  boundary
layer   (\citeauthor{holt03}  \citeyear{holt03};  \citeauthor{bear04b}
\citeyear{bear04b}), as  part of the GABLS initiative,  also used this
case (slightly modified).  Eleven different models with very different
SGS modeling options and different  grid-resolutions (from 1 m to 12.5
m)  were run  \cite{bear04b}.  In  this paper,  we also  simulate this
particular   case    with   our   newly    proposed   locally-averaged
scale-dependent dynamic (LASDD) SGS model. We compare our results with
theoretical  predictions by  Nieuwstadt \cite{nieu84a,nieu84b,nieu85},
field  observations,  as   well  as  with  well-established  empirical
formulations.

\subsection{Description of the Simulations}

The GABLS  LES intercomparison  case study is  described in  detail in
\citeauthor{bear04b}  (\citeyear{bear04b}).   The  boundary  layer  is
driven by  an imposed, uniform  geostrophic wind ($G = 8$ m s$^{-1}$),
with  a surface  cooling  rate of  $0.25$  K per  hour  and attains  a
quasi-steady state in $\sim$ 8-9  hours with a boundary layer depth of
$\sim 200$ m.  The initial  mean potential temperature is $265$ K upto
$100$ m with an overlying inversion of strength $0.01$ K m$^{-1}$. The
Coriolis parameter  is set  to $f_c =  1.39 \times  10^{-4}$ s$^{-1}$,
corresponding to latitude  $73^\circ$ N.  Our domain size  is: ($L_x =
L_y = L_z = 400$ m).  This domain is divided into: (1) $N_x \times N_y
\times N_z = 32 \times 32 \times 32$ nodes (i.e., $\Delta_x = \Delta_y
= \Delta_z = 12.5$  m); (2) $N_x \times N_y \times N_z  = 64 \times 64
\times 64$  nodes (i.e., $\Delta_x =  \Delta_y = \Delta_z  = 6.25$ m);
and (3)  $N_x \times N_y  \times N_z =  80 \times 80 \times  80$ nodes
(i.e., $\Delta_x = \Delta_y = \Delta_z  = 5$ m). One of the objectives
behind  these simulations was  to investigate  the sensitivity  of our
results to grid resolution.

A Galilean transformation is  used to weaken the stability constraints
on the time  step.  The grid moves with  ($U_{Gal},V_{Gal}) = (5.5,0)$
m/s  in the  $32^3$ and  $64^3$  nodes cases.  In the  case of  $80^3$
simulation, we  have used $(U_{Gal},V_{Gal})  = (5,0)$ m/s.   The time
steps ($\Delta t$) for our  $32^3$, $64^3$, and $80^3$ simulations are
0.4, 0.2 and 0.14 seconds, respectively.

\subsection{Description of the LES Code}

In  this  work, we  have  used  a modified  version  of  the LES  code
described       in       \citeauthor{albe99}      (\citeyear{albe99}),
\citeauthor{port00}   (\citeyear{port00}),   and   \citeauthor{port04}
(\citeyear{port04}).   The  salient  features  of  this  code  are  as
follows:
\begin{itemize}
\item  It  solves  the  filtered Navier-Stokes  equations  written  in
rotational form \cite{orsz74}.
\item Derivatives in the  horizontal directions are computed using the
Fourier   Collocation   method,   while   vertical   derivatives   are
approximated with second-order central differences \cite{canu88}.
\item Dealiasing of the nonlinear terms in Fourier space is done using
the $3/2$ rule \cite{canu88}.
\item Explicit second-order Adams-Bashforth time advancement scheme is
used \cite{canu88}.
\item Explicit  spectral cutoff filtering  is used. The  ratio between
the filter-width ($\Delta_f$) and  grid-spacing ($\Delta_g$) is set to
2.  Here $\alpha$ is taken to be equal to $\sqrt{2}$.
\item Only Coriolis terms involving horizontal wind are considered.
\item Forcing is imposed by Geostrophic wind.
\item Staggered vertical grid is used.
\end{itemize}

The lower boundary condition  is based on the Monin-Obukhov similarity
theory with  a surface roughness length  $z_\circ$.  The instantaneous
components of  surface shear stresses $\tau_{xz}$  and $\tau_{yz}$ are
represented as  functions of  the resolved velocities  $\tilde{u}$ and
$\tilde{v}$, at the grid point immediately above the surface (i.e., at
a height of $z = \Delta_z/2$ in our case):
\begin{equation}
\tau_{xz} = -u_*^2\left[\frac{\tilde{u}(z)}{U(z)}\right]
\end{equation}
\begin{equation}
\tau_{yz} = -u_*^2\left[\frac{\tilde{v}(z)}{U(z)}\right],
\end{equation}
where $u_*$ is the friction  velocity, which is computed from the mean
horizontal   wind speed    $U(z)   =   \langle   (\tilde{u}^2   +
\tilde{v}^2)^{1/2} \rangle$  at the first  vertical model level  ($z =
\Delta_z/2$) as follows:
\begin{equation}
u_* = \frac{U(z)\kappa}{\log(\frac{z}{z_\circ})+b_m \frac{z}{L}}.
\end{equation}
In a similar manner, the surface heat flux is computed as:
\begin{equation}
\left\langle{w\theta}_s \right\rangle =    
\frac{u_*    \kappa   \left[\theta_s    -
\Theta(z)\right]}{\log(\frac{z}{z_\circ})+b_h \frac{z}{L}},
\end{equation}
where $\theta_s$  and $\Theta(z)$  denote the surface  temperature and
the  mean resolved  potential temperature  at the  first  model level,
respectively.    Following   the    recommendations   of   the   GABLS
intercomparison study, the  constants $b_m$ and $b_h$ were  set to 4.8
and 7.8, respectively.

The upper  boundary consists of  a zero stress condition,  whereas the
lateral  boundary condition  assumes periodicity.  A  Rayleigh damping
layer at 300 m is used following the GABLS case description.
 
\subsection{Results}

In this section  we report the results of  our tuning-free simulations
and attempt  to evaluate them against the  theoretical predictions and
well  established observations-based  formulations.  Our results  show
clear improvements over most of  the traditional models in the surface
layer.  We would  like  to emphasize  that  in the  surface layer  the
relative contribution  of the SGS  to the overall turbulent  fluxes is
very  large.  This is  also  the  location  where gradients  are  much
stronger.   Hence, simulation  results  become very  sensitive to  SGS
formulations  near  the  ground.    The  comparisons  we  perform  are
extensive and address: (1) temporal evolution of simulated statistics,
(2) various   first  order  statistics   of  turbulent   velocity  and
temperature fields, (3) second  order statistics of turbulent velocity
and  temperature  fields,   and  (4)  characteristics  of  dynamically
estimated SGS coefficients.  Details of these comparisons are provided
below.

The  boundary  layer height  ($H$),  Obukhov  length  ($L$) and  other
characteristics  of  the  simulated  SBLs using  the  locally-averaged
scale-dependent  dynamic SGS model  (averaged over  the final  hour of
simulation)    are   given    in   Table    \ref{TBasic}.    Following
\citeauthor{koso00}   (\citeyear{koso00})   and   \citeauthor{bear04b}
(\citeyear{bear04b}), the  boundary layer  height ($H$) is  defined as
$(1/0.95)$ times the height where  the mean local stress falls to five
percent of  its surface value.  From  this table, it  is apparent that
the simulated  (bulk) boundary layer parameters  are quite insensitive
to the grid resolution.  In  LES this behavior is always desirable and
its existence is usually attributed to the strength of a SGS model.

\begin{table}
\caption{Basic characteristics of the simulated SBLs during the last
hour of simulation.}\label{TBasic}
\begin{tabular}{lcccc} \hline \hline
Grid Points & $H$ (m) & $L$ (m) & $u_*$ (ms$^{-1}$) & $\theta_*$ (K)
\\ \hline 
$32\times 32\times 32$ & 205 & 113 & 0.283 & 0.047 \\
$64\times 64\times 64$ & 185 & 114 & 0.276 & 0.045   \\
$80\times 80\times 80$ & 192 & 122 & 0.285 & 0.045 \\ \hline
\end{tabular} 
\end{table}

\subsubsection{Temporal Evolution}

\begin{figure}[tb]
\figbox*{}{}  {\epsfxsize=3.25 in \epsfbox{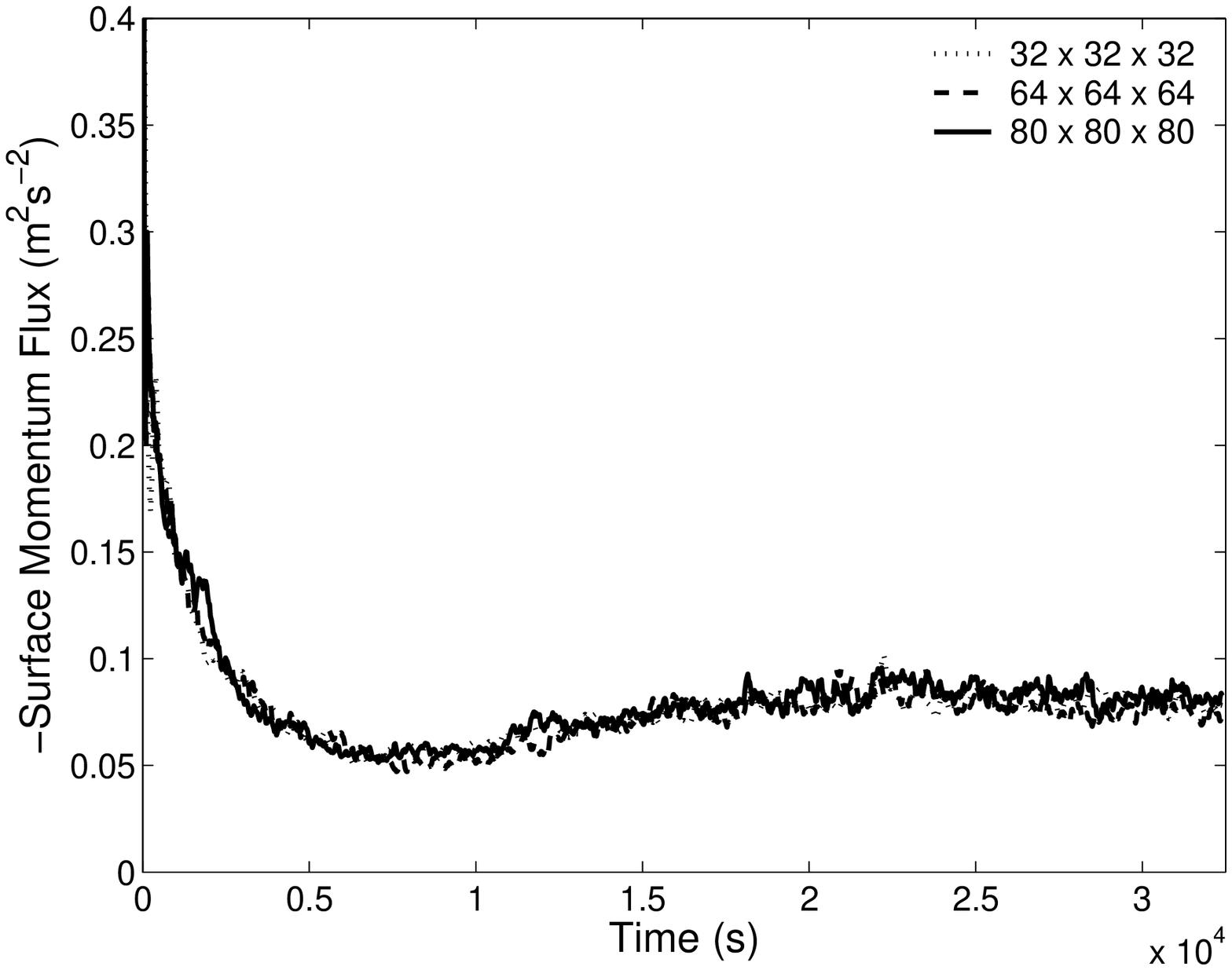}}
\figbox*{}{}  {\epsfxsize=3.25 in \epsfbox{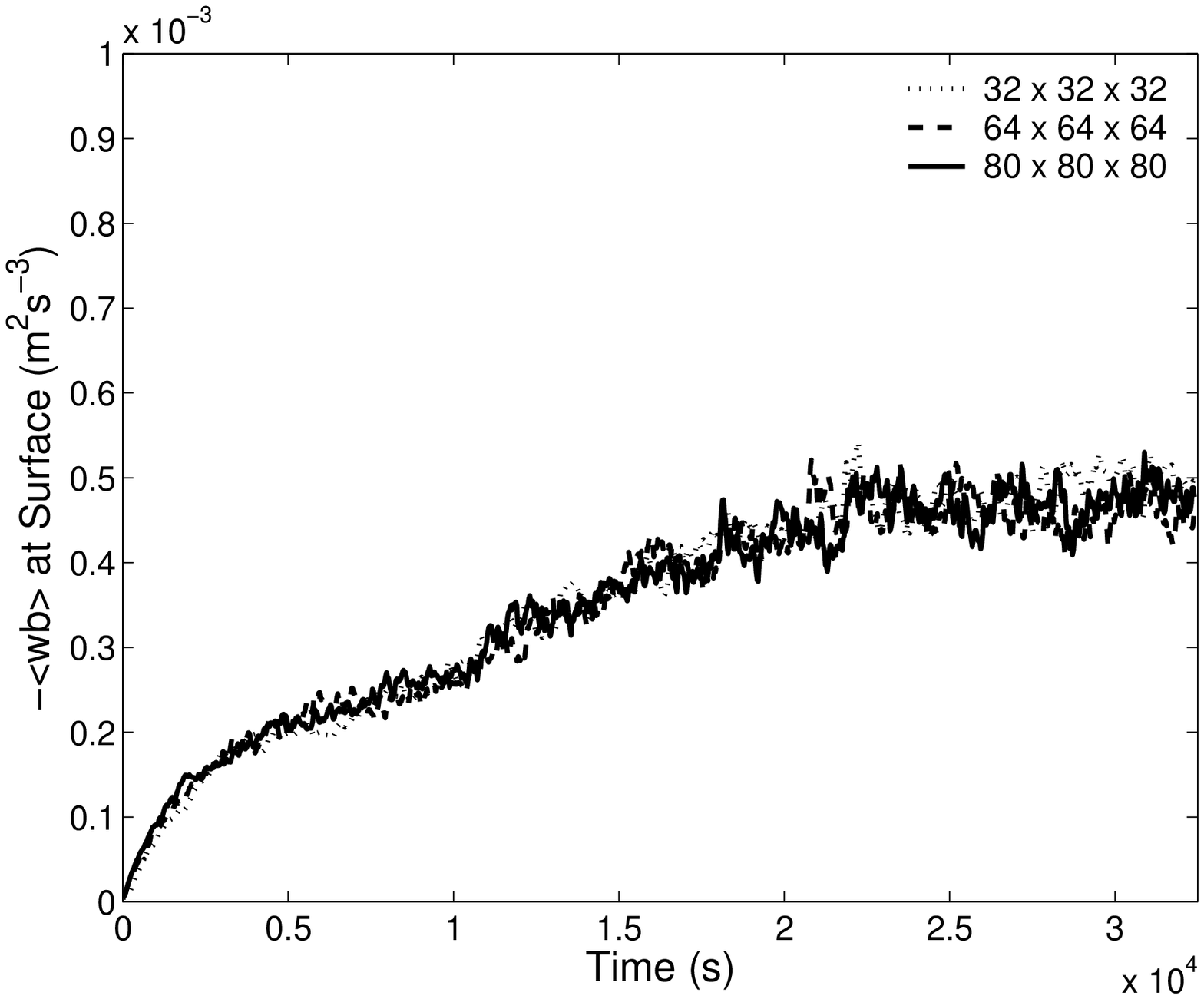}}
\caption{\label{FigStableMomHeatFluxTime}   Time  series   of  surface
momentum flux (top) and surface buoyancy flux (bottom).}
\end{figure}

\begin{figure}[tb]
\figbox*{}{} {\epsfxsize=3.25 in  \epsfbox{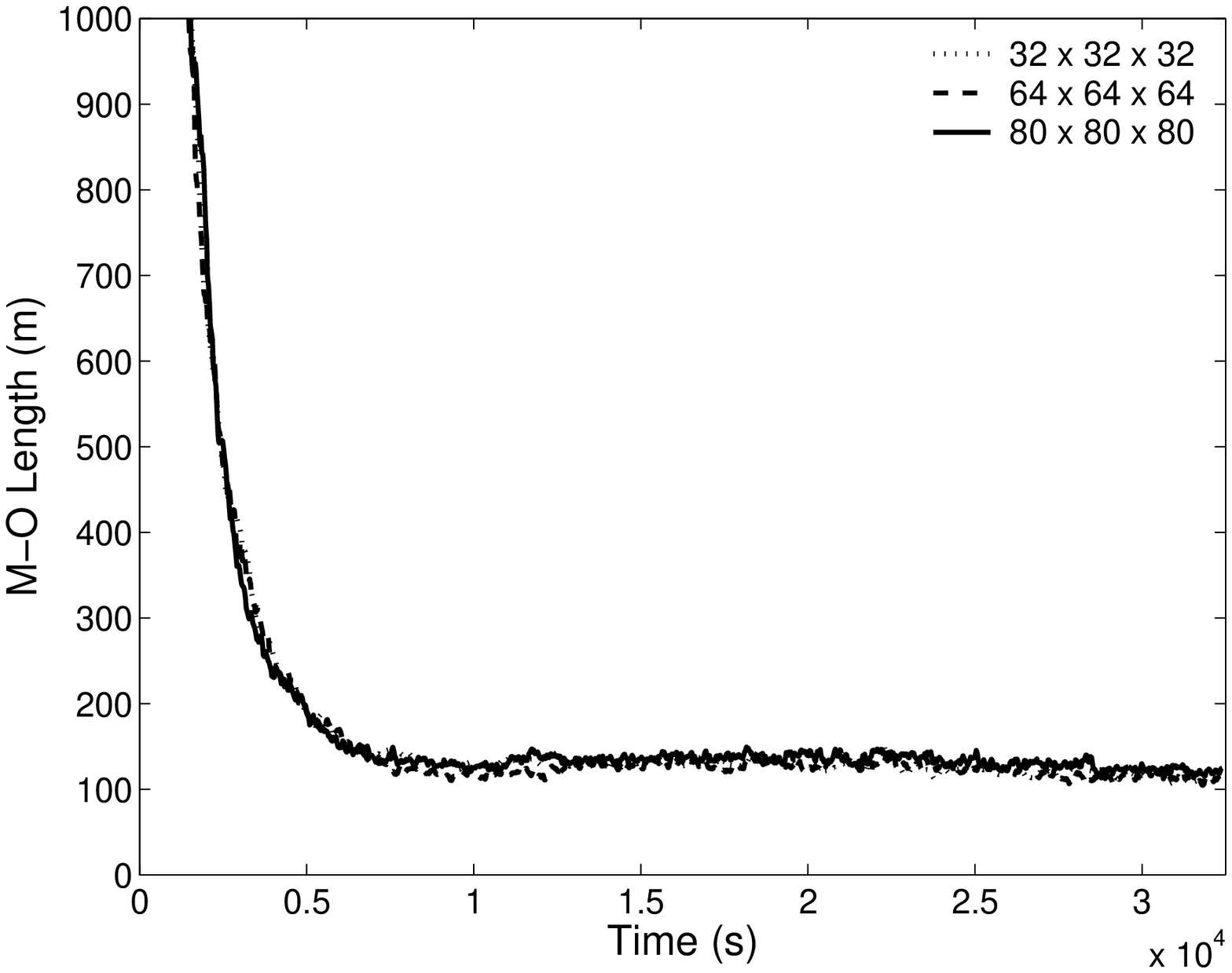}}
\figbox*{}{} {\epsfxsize=3.25 in  \epsfbox{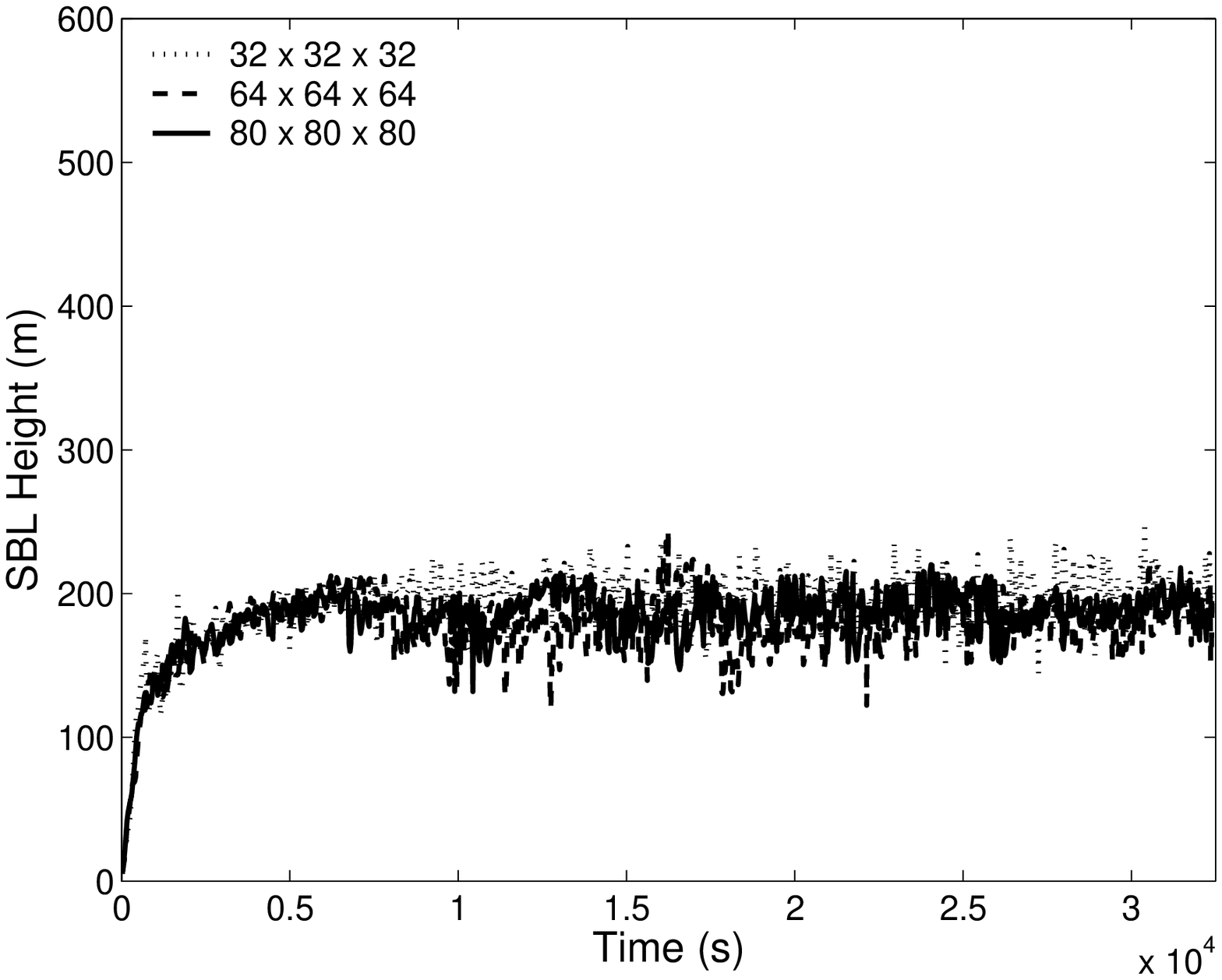}}
\caption{\label{FigStableMOHTime} Time series  of Obukhov length (top)
and boundary layer height (bottom).}
\end{figure}

In  Figures \ref{FigStableMomHeatFluxTime}  and \ref{FigStableMOHTime}
the  time series  of  surface momentum  flux,  surface buoyancy  flux,
Obukhov length  ($L$) and boundary  layer height ($H$) are  shown. The
surface  momentum  flux reaches  quasi-equilibrium  after  4 hours  of
simulation.  On the  other hand the Obukhov length  and boundary layer
height  equilibrate well  before  2 hours  of  simulation.  Since  the
surface  boundary condition  is  prescribed by  a  fixed cooling  rate
rather than a fixed flux,  it is anticipated that the surface buoyancy
flux will keep on evolving with time.

\subsubsection{First-order Statistics}

\begin{figure}[tb]
\figbox*{}{}     {\epsfxsize=3.25     in    \epsfbox{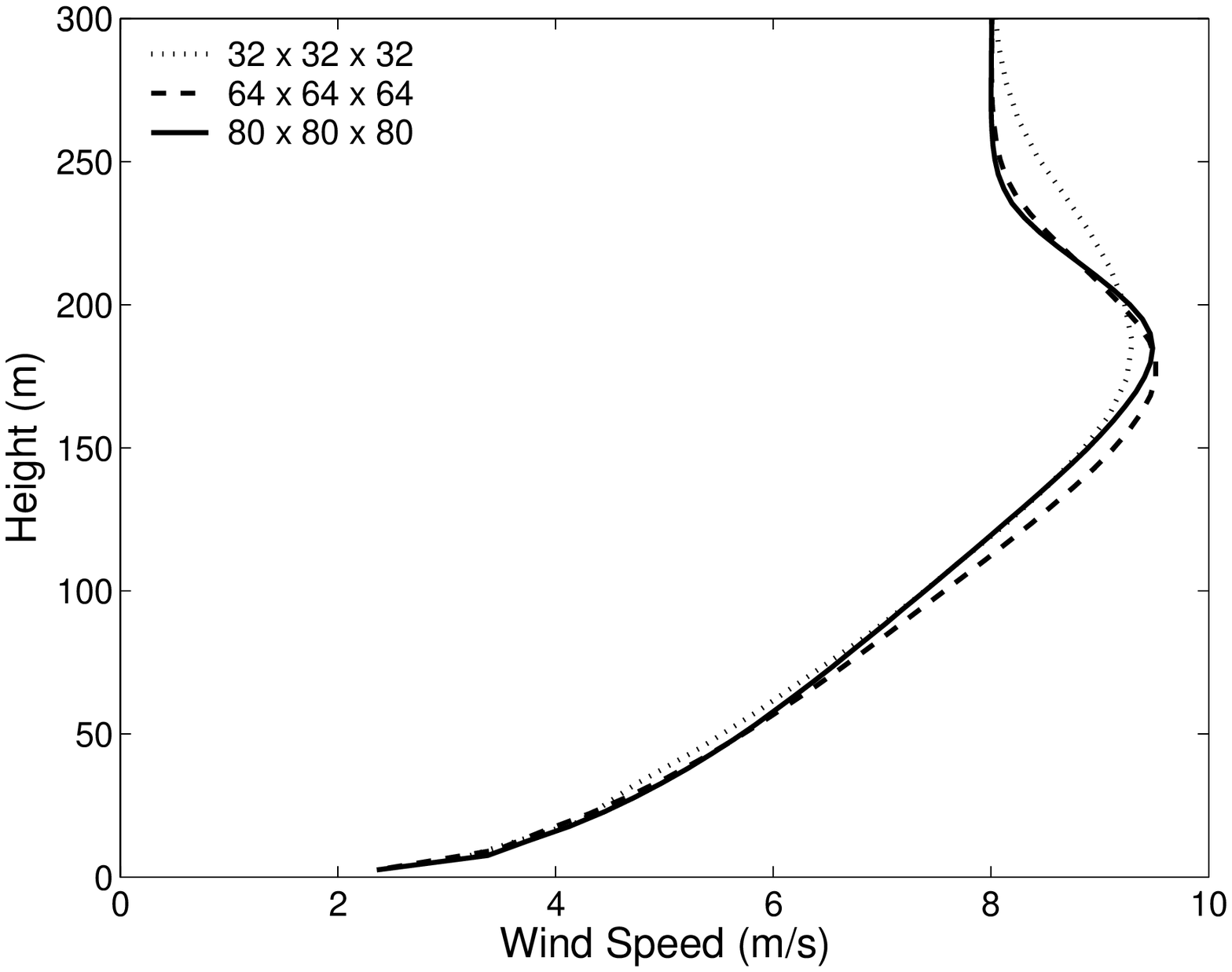}}
\figbox*{}{} {\epsfxsize=3.25 in \epsfbox{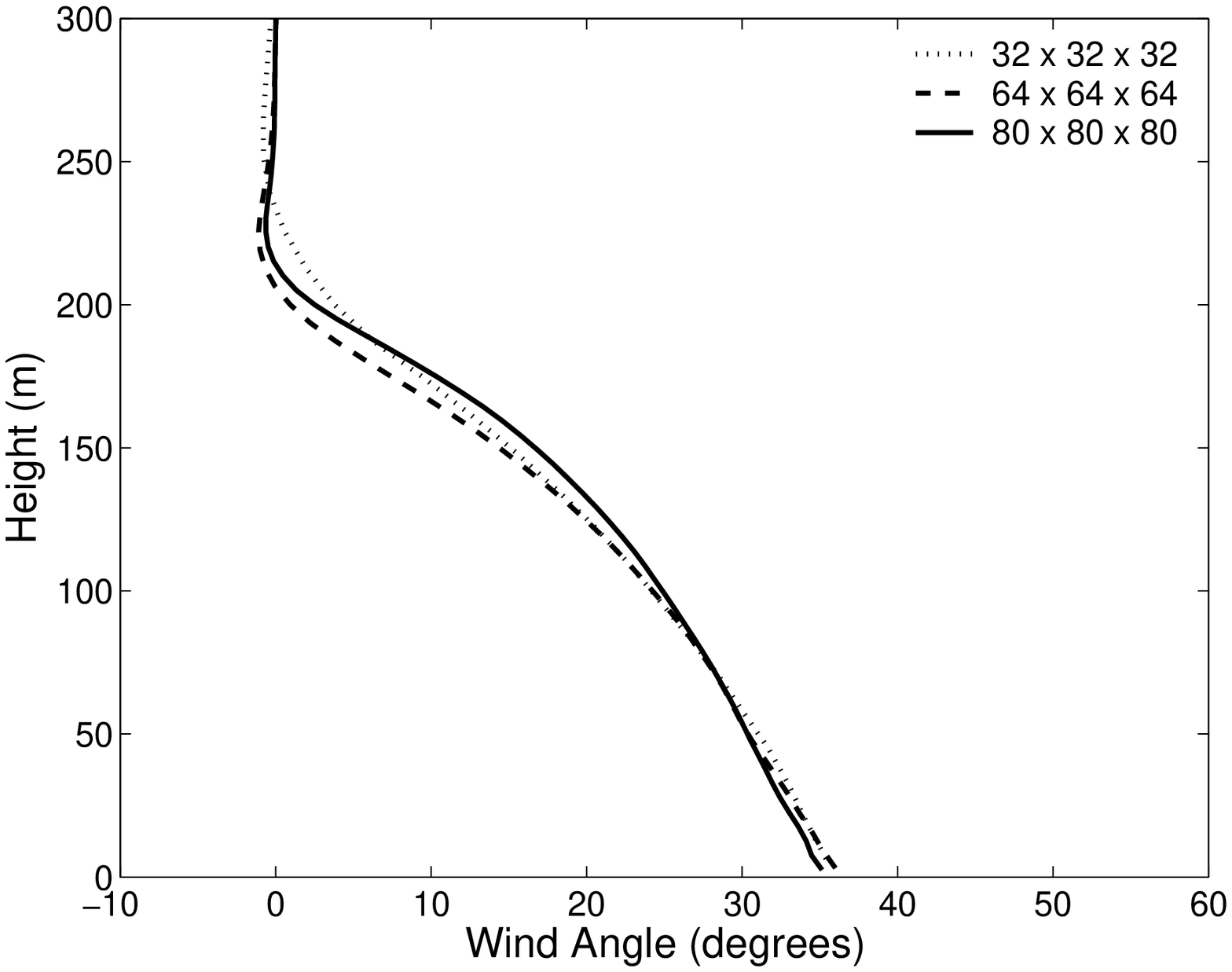}}
\caption{\label{FigStableMX}  Mean  wind speed  (top)  and wind  angle
(bottom) profiles. These profiles are  averaged over the last one hour
of simulation.}
\end{figure}

\begin{figure}[tb]
\figbox*{}{}{\epsfxsize=3.25 in  \epsfbox{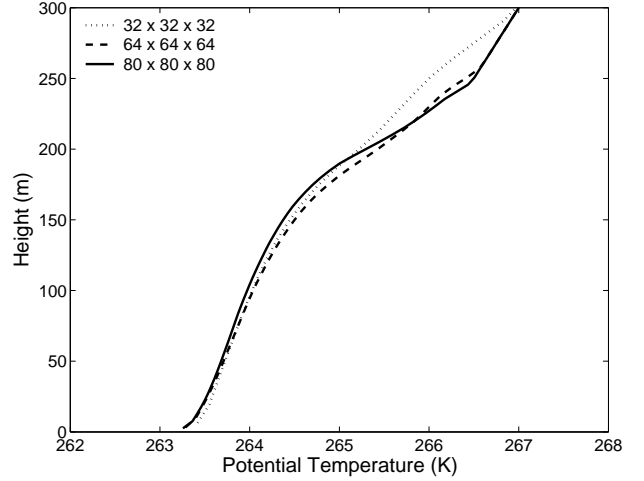}}
\caption{\label{FigStableT} Mean  temperature profiles. These profiles
are averaged over the last one hour of simulation.}
\end{figure}

The mean profiles of wind  speed, wind angle and potential temperature
averaged over the  final hour (8-9 hours) of  simulation, are shown in
Figures \ref{FigStableMX} and \ref{FigStableT}.  The super-geostrophic
nocturnal jet  near the  top of the  boundary layer, is  in accordance
with Nieuwstadt's  theoretical model for  `stationary' stable boundary
layers [see  Equation 17 of  \citeauthor{nieu85} (\citeyear{nieu85})].
However,  the  angle  between  the  surface  wind  direction  and  the
geostrophic wind simulated  by our LES is $\sim  35$ degrees.  This is
much  smaller  than  Nieuwstadt's   prediction  of  60  degrees.   The
second-order closure  model of \citeauthor{bros78} (\citeyear{bros78})
also predicts a value of  $\sim 58$ degrees.  In contrast, the results
from   the   GABLS   study  \cite{bear04b}   and   \citeauthor{koso00}
(\citeyear{koso00}) are in agreement with our results.

Nieuwstadt  also  derived   the  following  mean  temperature  profile
[Equation 21 of \citeauthor{nieu85} (\citeyear{nieu85})]:
\begin{equation}
\label{NieuwTemp}
\frac{\Theta-\theta_s}{\theta_*}         =         -\frac{Ri_g}{\kappa
Ri_f^2}\frac{H}{L} \ln\left(1-\frac{z}{H}\right),
\end{equation}
where $Ri_f$  and $Ri_g$ denote  the flux and the  gradient Richardson
numbers,                    respectively.                    $\theta_*
\left(=-\frac{\left\langle{w\theta}\right\rangle}{u_*}\right)$  
signifies  the surface
layer  temperature scale.  Equation  \ref{NieuwTemp} implies  that the
temperature    profile     exhibits    positive    curvature    ($\sim
\partial^2\Theta/\partial  z^2$), which is  clearly visible  in Figure
\ref{FigStableT}.

\begin{figure}[tb]
\figbox*{}{} {\epsfxsize=3.25 in  \epsfbox{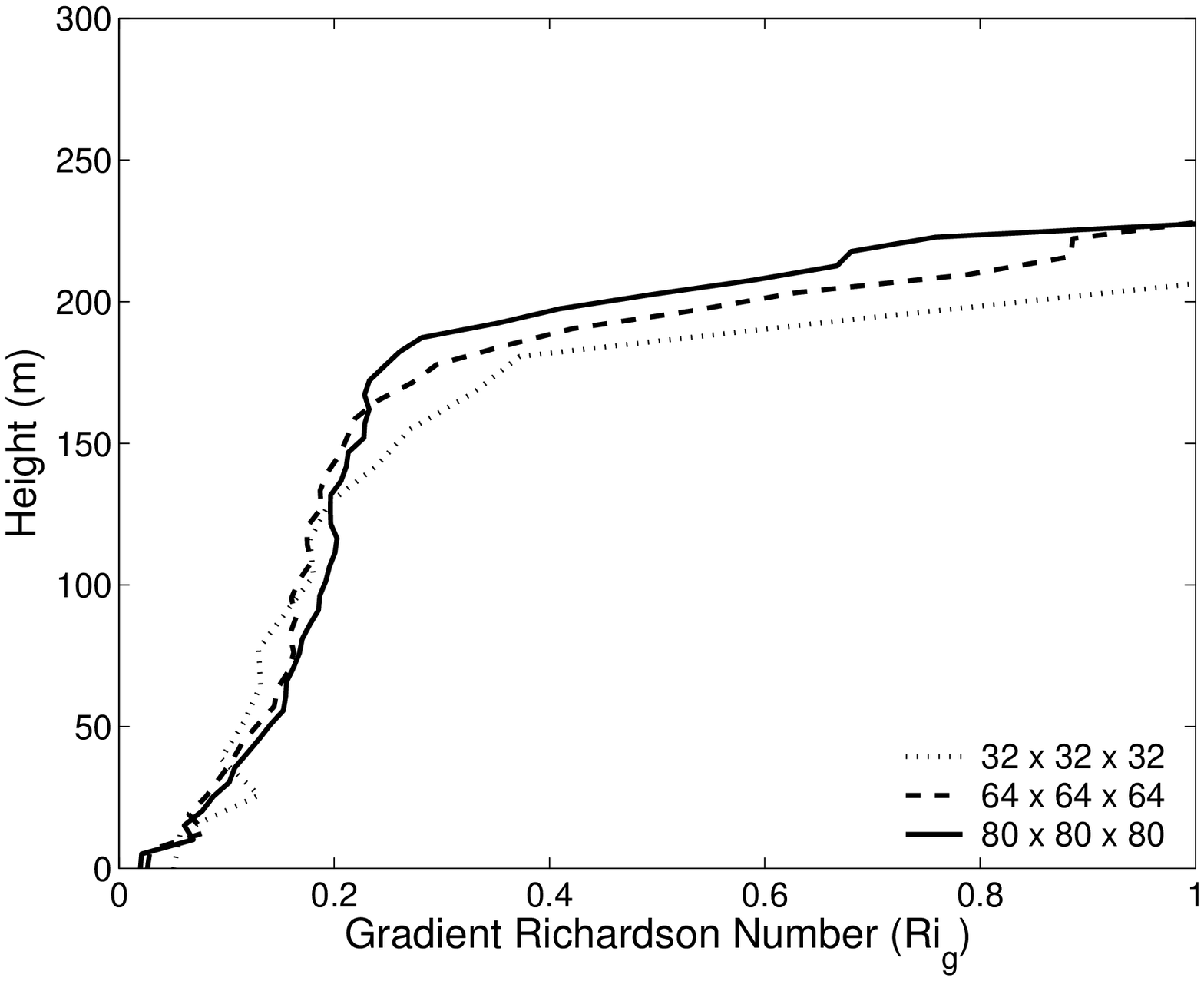}}
\figbox*{}{} {\epsfxsize=3.25 in  \epsfbox{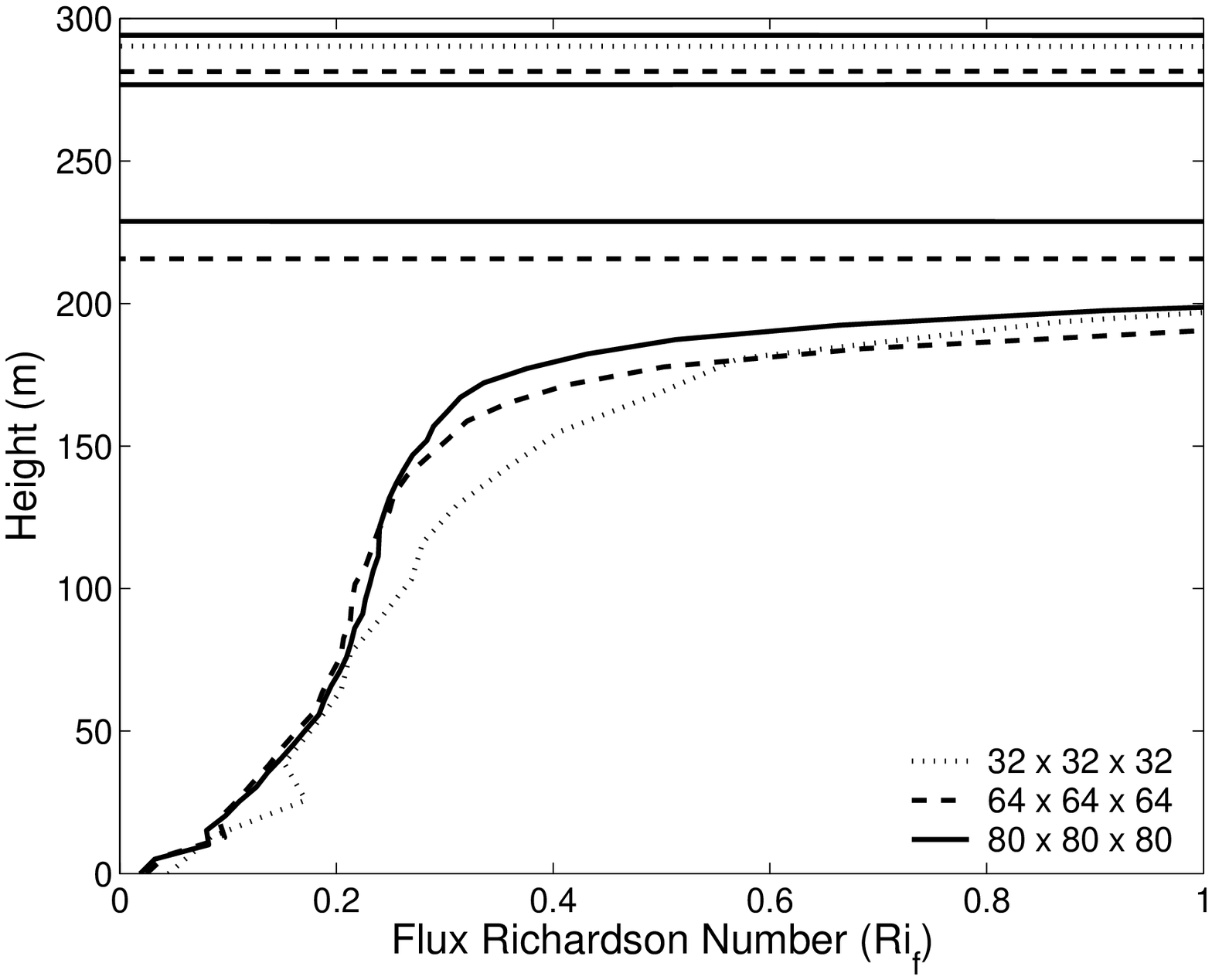}}
\caption{\label{FigStableRi}  Mean  profiles  of  gradient  Richardson
number (top) and flux  Richardson number (bottom).  These profiles are
averaged over the last one hour of simulation.}
\end{figure}

We would like to point out that Nieuwstadt's analytical model is based
on the hypothesis that the gradient Richardson number ($Ri_g$) and the
flux Richardson  number ($Ri_f$) are  constant with height  inside the
stable boundary  layer.  Nieuwstadt  was aware of  the fact  that this
hypothesis  does not hold  for the  lower part  of the  boundary layer
\cite{nieu85}.  In fact, $Ri_g$ and  $Ri_f$ should go to zero near the
surface  \cite{nieu85}, as  can be  seen from  our simulations (Figure
\ref{FigStableRi}).   The violation  of  the basic  assumption in  the
proximity of the land surface  might explain some of the discrepancies
between the LES results and Nieuwstadt's predictions.

The Richardson numbers represent the  ratio of the amount of turbulent
kinetic energy  (TKE) destroyed by  buoyancy forces to the  amount of
TKE generated by  wind shear \cite{stul88}.  The values  of $Ri_f$ are
consistently  higher than  the corresponding  $Ri_g$ values,  which is
expected  (see below).  In the  interior part  of the  boundary layer,
$Ri_g$  is  more  or  less  constant  ($\sim  0.2$),  in  accord  with
Nieuwstadt's assumption.  However,  $Ri_f$ increases monotonically and
is  higher than  0.2 in  the upper  part of  the boundary  layer.  The
magnitudes of both these  Richardson numbers increase sharply near the
top of  the boundary  layer and  become more than  1 in  the inversion
layer.

It is straightforward to show that the ratio between $Ri_g$ and $Ri_f$
is the turbulent Prandtl number ($Pr_t$) \cite{derb99,howe99}:
\begin{equation}
Pr_t = \frac{K_M}{K_H} = \frac{Ri_g}{Ri_f},
\end{equation} 
where $K_M$  and $K_H$ represent  eddy diffusivities for  momentum and
heat  flux,  respectively. The  dependence  of  $Pr_t$ on  atmospheric
stability  is not  strong \cite{derb99,howe99}.   Inside  the boundary
layer (up to $\sim$ 150 m),  (almost) all our simulated  results yield
$\frac{Ri_g}{Ri_f}  =  Pr_t \sim  0.7$  (not  shown  here).  Based  on
phenomenological    theories    of   turbulence    \citeauthor{town76}
(\citeyear{town76})  and \citeauthor{yakh86}  (\citeyear{yakh86}) also
derived $Pr_t = 0.7$.  However,  in the surface layer our results show
that the  values of  $Pr_t$ increase to  $\sim 1$. This  is consistent
with    `Microfronts'   field    experimental    data   analyzed    by
\citeauthor{howe99}  (\citeyear{howe99}). They  found on  average, the
estimates of $Pr_t$  at 3 m level  are higher than at the  10 m level,
indicating that the relative efficiency of turbulent momentum transfer
with respect to  heat transfer increases in the  proximity of the land
surface \cite{howe99}.

\begin{figure}[tb]
\figbox*{}{} {\epsfxsize=3.25 in \epsfbox{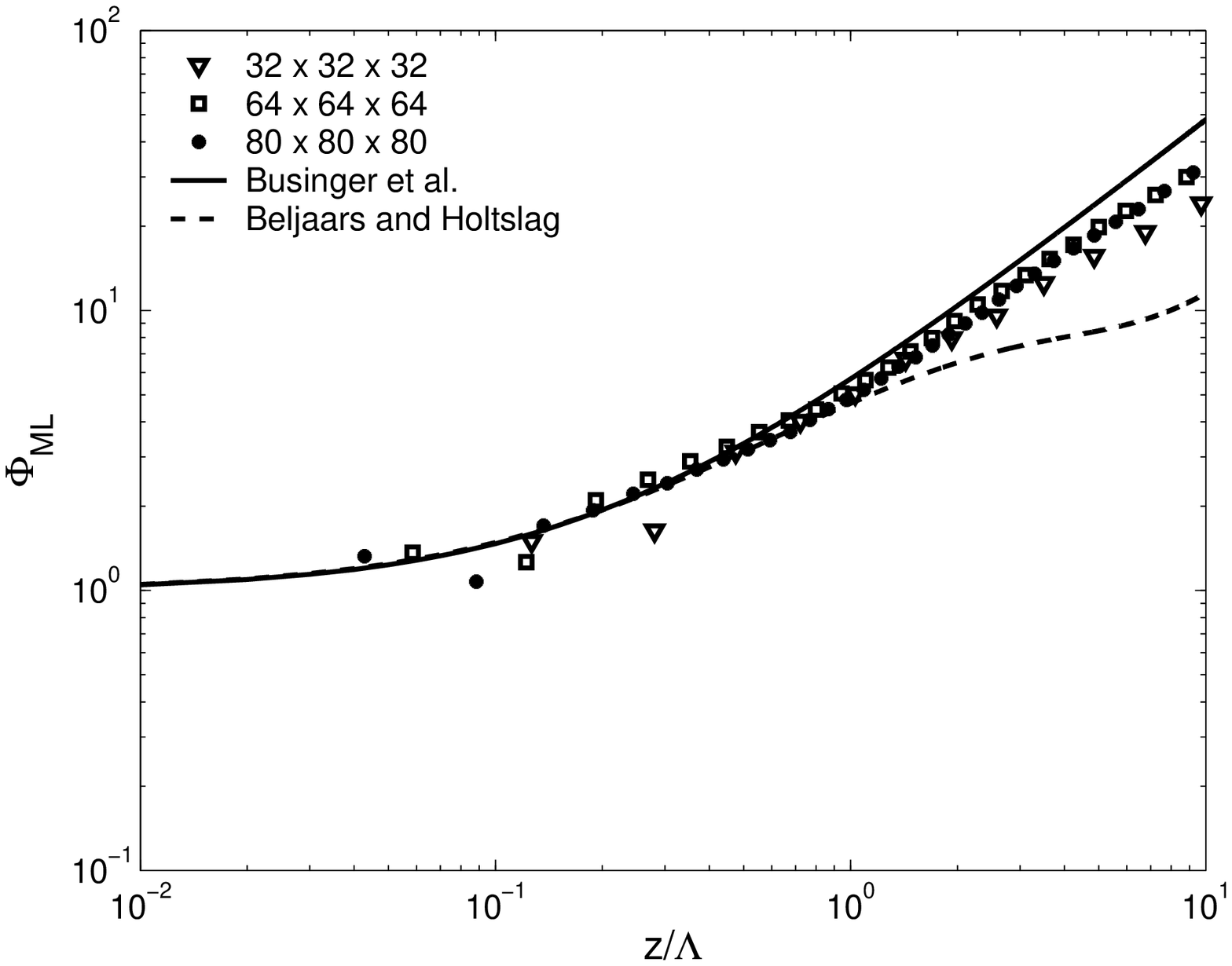}}
\figbox*{}{} {\epsfxsize=3.25 in \epsfbox{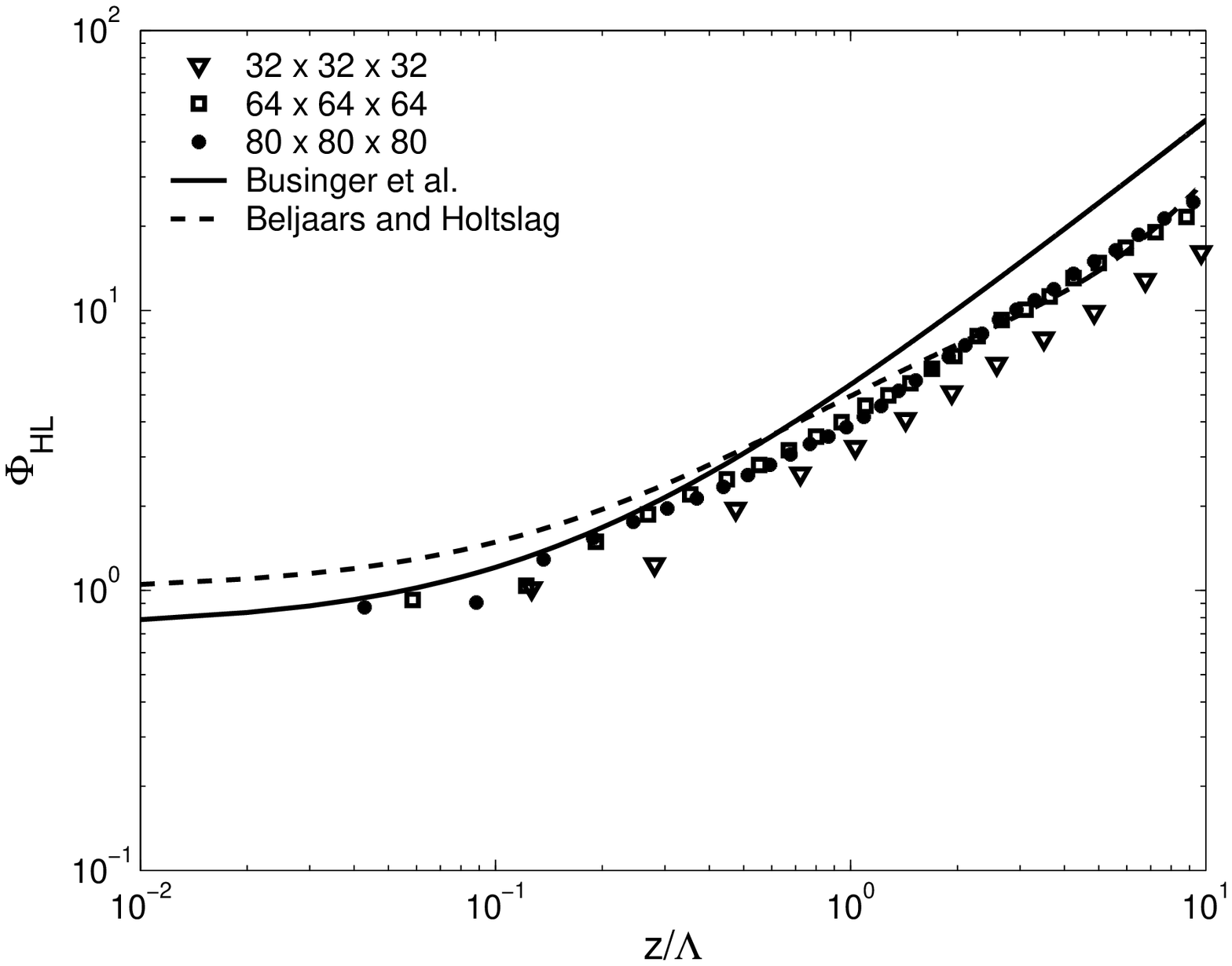}}
\caption{\label{FigStablePhiL}    Locally    computed   nondimensional
gradients of velocity (top) and temperature (bottom) against the local
stability parameter. These statistics are computed during the last one
hour of simulation. The field-observations-based formulations given by
\citeauthor{busi71}    (\citeyear{busi71})   and   \citeauthor{belj91}
(\citeyear{belj91}) are also shown.}
\end{figure}

In SBL  simulations, one can  test the performance  of a SGS  model by
plotting a local nondimensional shear:
\begin{equation}
\Phi_{ML}   =   \frac{\kappa   z}{u_{*L}}   \sqrt{\left(\frac{\partial
U}{\partial   z}    \right)^2   +   \left(\frac{\partial   V}{\partial
z}\right)^2}
\end{equation}
and nondimensional temperature gradient:
\begin{equation}
\Phi_{HL}     =     \frac{\kappa    z}{\theta_{*L}}     \frac{\partial
\Theta}{\partial z}
\end{equation}
as a function of local stability parameter ($z/\Lambda$) and comparing
with  field-observations-based formulations.  Here,  $\Lambda$ denotes
the local  Obukhov length.   In this work,  a subscript `$_L$'  on the
turbulence  quantities  (e.g.,  $u_{*L}$)  will  be  used  to  specify
evalutation  using local  turbulence quantities  -  otherwise, surface
values are implied.  Recently, \citeauthor{mahr03} (\citeyear{mahr03})
called  this type  of  similarity theory  `hybrid similarity  theory',
since it approaches Monin-Obukhov similarity as $z$ decreases and also
conforms  to  z-less stratification  as  $z  \rightarrow \infty$.   In
Figure  \ref{FigStablePhiL} we  plot  the `hybrid'  nondimensionalized
gradients    and    compare   them    with    the   formulations    by
\citeauthor{busi71} (\citeyear{busi71}):
\begin{equation}
\Phi_{ML} = 1+4.7\frac{z}{\Lambda}
\end{equation}
\begin{equation}
\Phi_{HL} = 0.74+4.7\frac{z}{\Lambda},
\end{equation}
and by \citeauthor{belj91} (\citeyear{belj91}):
\begin{equation}
\Phi_{ML}     =    1+\frac{z}{\Lambda}\left[a+be^{-d\frac{z}{\Lambda}}
\left(1+c-d\frac{z}{\Lambda}\right) \right]
\end{equation}
\begin{equation}
\Phi_{HL}        =       1+\frac{z}{\Lambda}\left[a\left(1+\frac{2}{3}
\frac{az}{\Lambda}\right)^{1/2}+be^{-d\frac{z}{\Lambda}}
\left(1+c-d\frac{z}{\Lambda}\right) \right],
\end{equation}
where the suggested values of the coefficients are \cite{belj91}: $a =
1,  b =  2/3,  c  = 5$  and  $d =  0.35$.   Interestingly, both  these
simulated gradients  plotted against $z/\Lambda$  show slopes slightly
smaller than the widely used Businger et al.'s formulations.  Based on
CASES99      field     observations      data,     \citeauthor{mahr03}
(\citeyear{mahr03}) found  a slope of  $3.7$ [in contrast to  $4.7$ as
proposed by \citeauthor{busi71}  (\citeyear{busi71})], which also fits
our  LES   results  remarkably   well.   Previous  studies,   such  as
\citeauthor{belj91}  (\citeyear{belj91}) also  found  that $\Phi_{ML}$
and $\Phi_{HL}$  increase slower  than Businger et  al.'s formulations
and   they   proposed   the   aforementioned   nonlinear   formulation
\cite{belj91}.

\begin{figure}[tb]
\figbox*{}{} {\epsfxsize=3.25 in \epsfbox{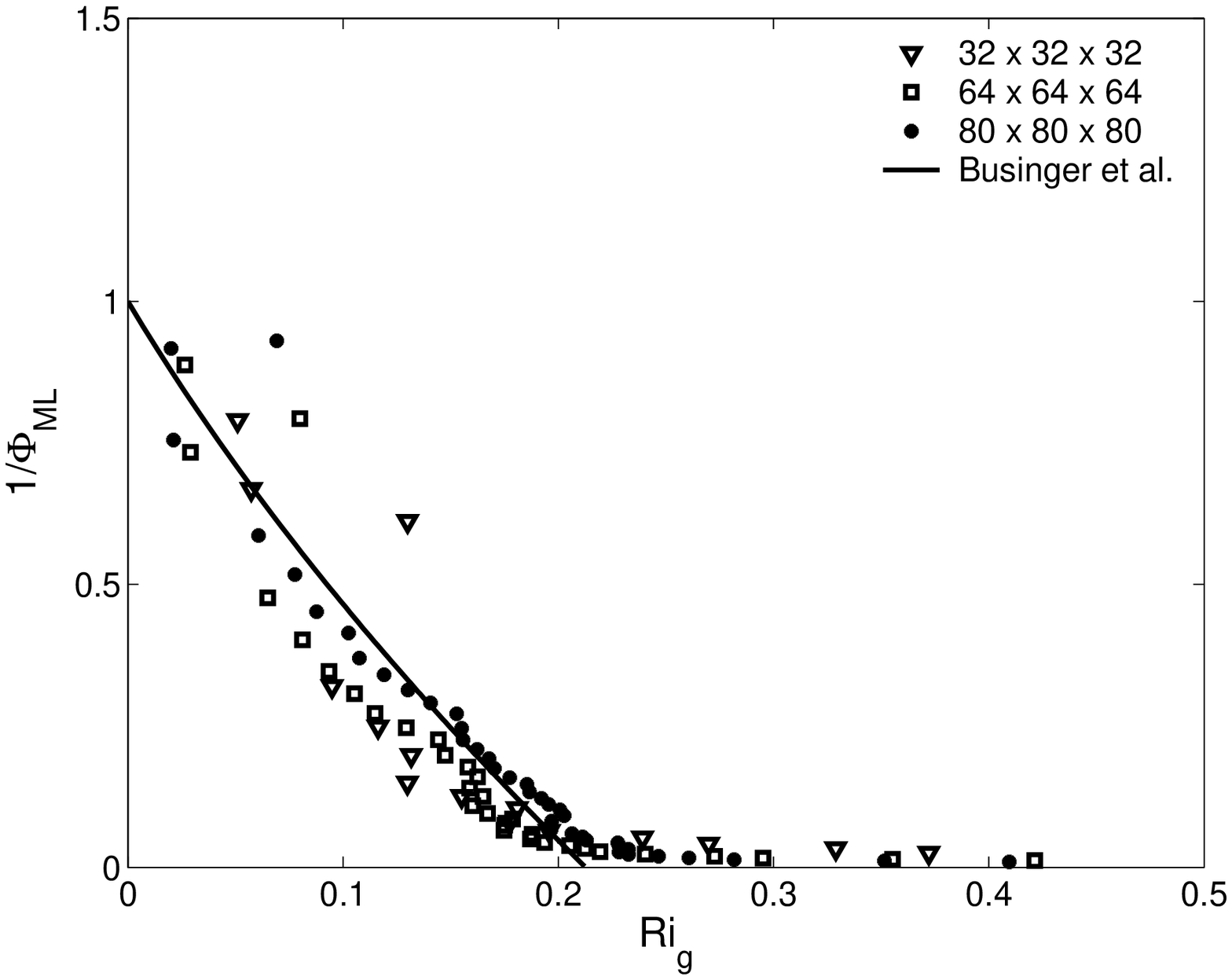}}
\figbox*{}{} {\epsfxsize=3.25 in \epsfbox{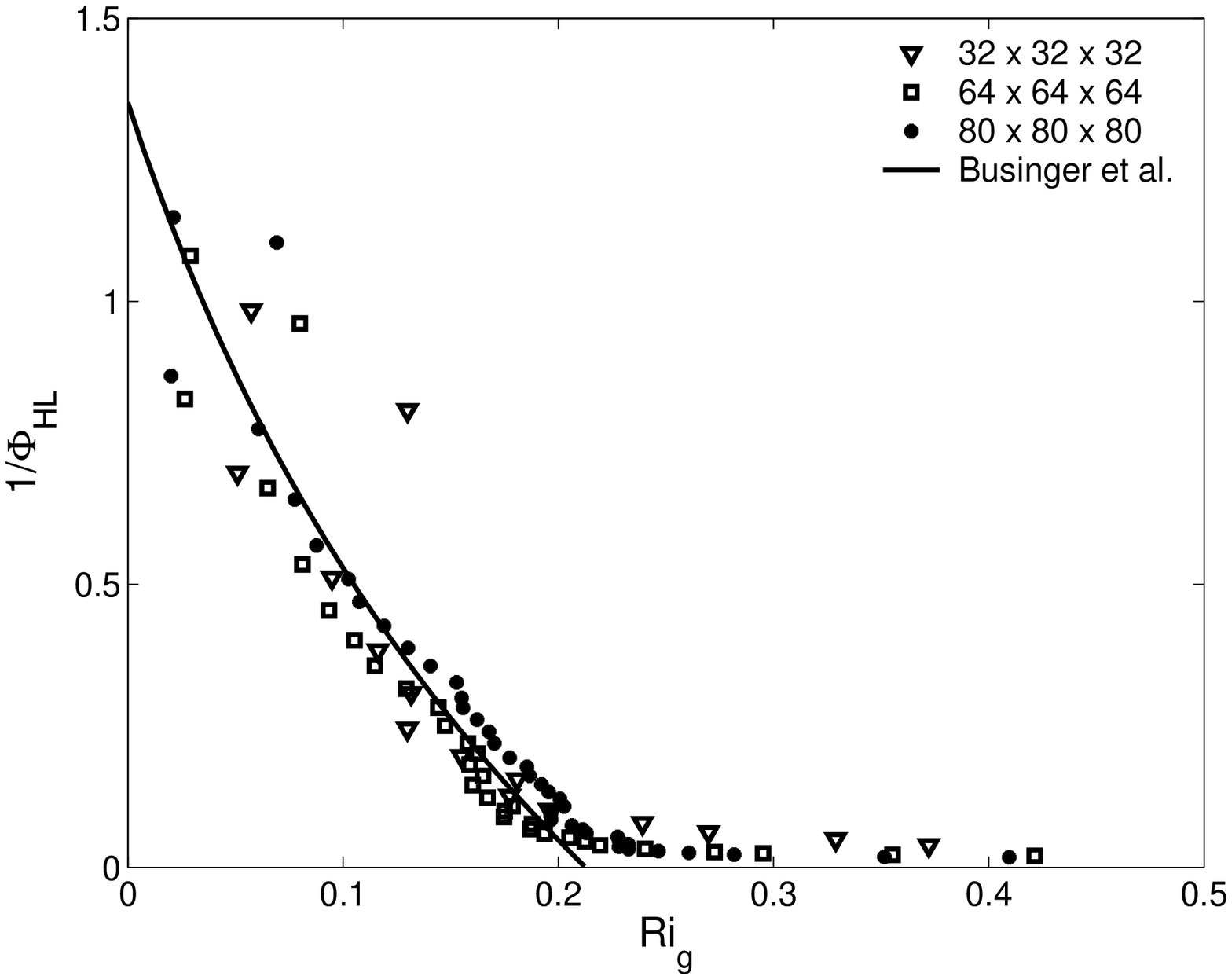}}
\caption{\label{FigStablePhiRi}    Locally   computed   nondimensional
gradients  of velocity  (top) and  temperature (bottom)  against local
gradient Richardson  number. These statistics are  computed during the
last   one   hour   of   simulation.   The   formulations   given   by
\citeauthor{busi71} (\citeyear{busi71}) are also shown.}
\end{figure}

There exists  another representation for  the nondimensional gradients
in  terms  of  local  gradient  Richardson  number  ($Ri_g$).   Figure
\ref{FigStablePhiRi} once  again shows that the  agreement between our
LES results  and established formulations are  quite satisfactory.  In
the  literature,  usually  the  critical  gradient  Richardson  number
($Ri_{gc}$) is  considered to  be around 0.25.   For $Ri_{gc}  > 0.25$
turbulence   is  very  weak   (\citeauthor{stul88}  \citeyear{stul88};
\citeauthor{brow94}  \citeyear{brow94}), which  is also  noticeable in
Figure \ref{FigStablePhiRi}.

\subsubsection{Second-order Statistics}

\begin{figure}[tb]
\figbox*{}{} {\epsfxsize=3.25 in  \epsfbox{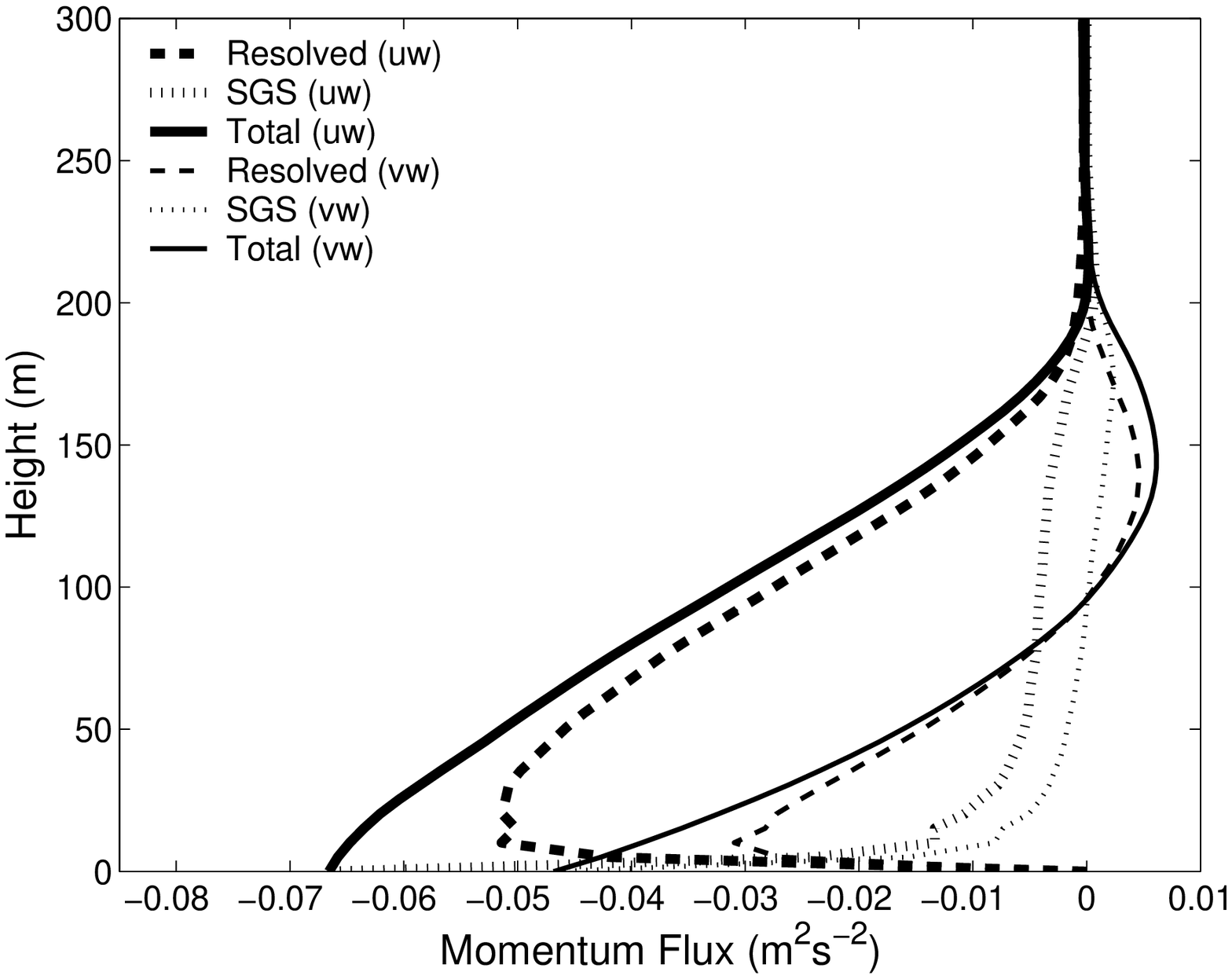}}
\figbox*{}{} {\epsfxsize=3.25 in  \epsfbox{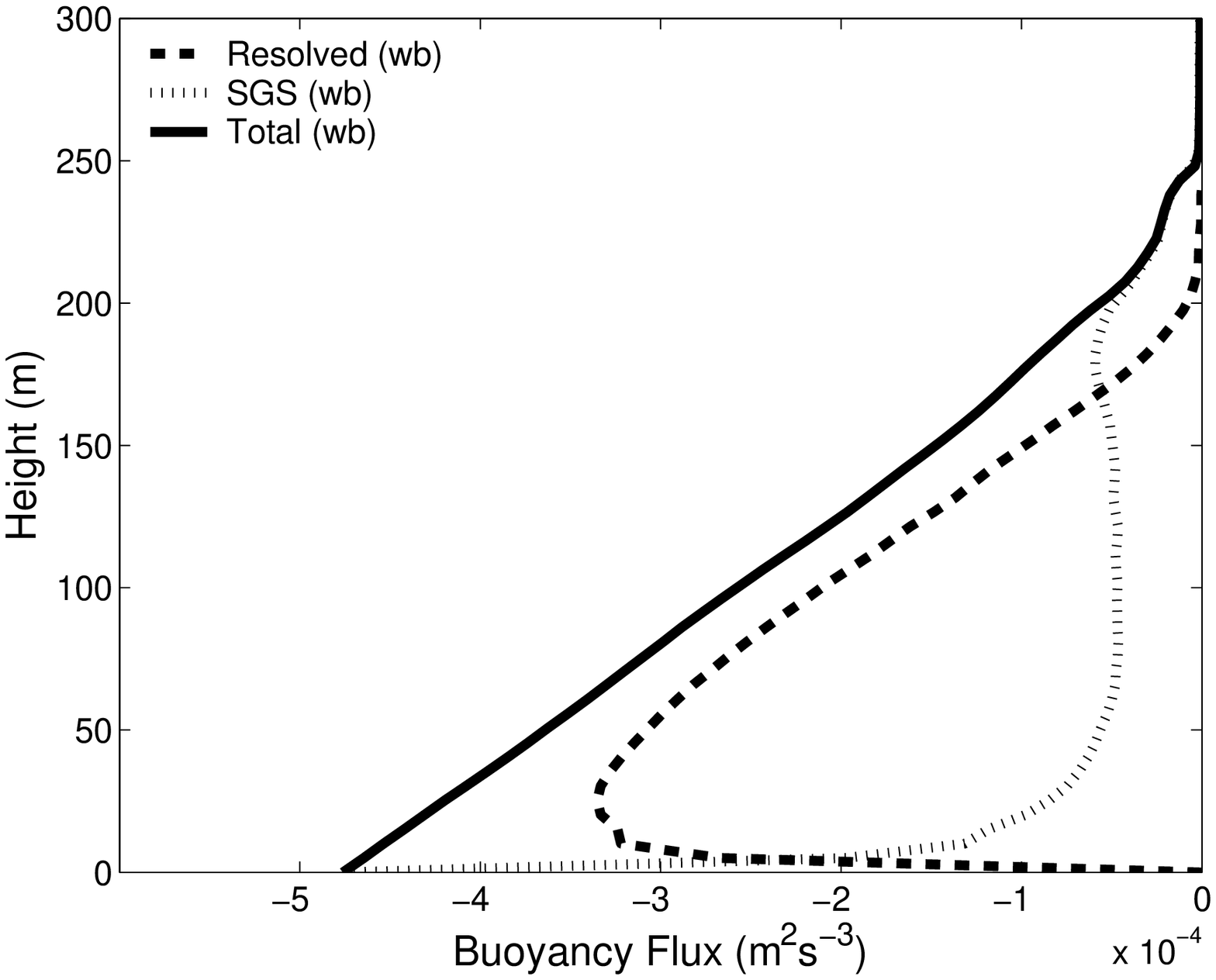}}
\caption{\label{FigStableFlux} Mean  momentum flux (top)  and buoyancy
flux (bottom)  profiles corresponding  to the $80^3$  simulation using
the locally-averaged scale-dependent  dynamic (LASDD) SGS model. These
profiles are averaged over the last one hour of simulation.}
\end{figure}

The mean profiles of vertical momentum flux and buoyancy flux averaged
over  the final hour  (8-9 hours)  of simulation  are given  in Figure
\ref{FigStableFlux}. The dashed lines  show the resolved flux, and the
dotted  lines denote the  SGS contribution  to the  flux. As  would be
anticipated, near the ground the  SGS contribution is much larger than
its resolved  counterpart.  In the GABLS  intercomparison study, there
is a significant  spread between the total momentum  and buoyancy flux
profiles simulated with different models. In particular at the surface
the  mean  momentum  and  buoyancy  fluxes range  from  0.06  to  0.08
m$^2$s$^{-2}$  and  -3.5   to  -5.5  $\times  10^{-4}$  m$^2$s$^{-3}$,
respectively \cite{bear04b}. Our simulated  results also fall in these
ranges.

\begin{figure}[tb]
\figbox*{}{} {\epsfxsize=3.25 in \epsfbox{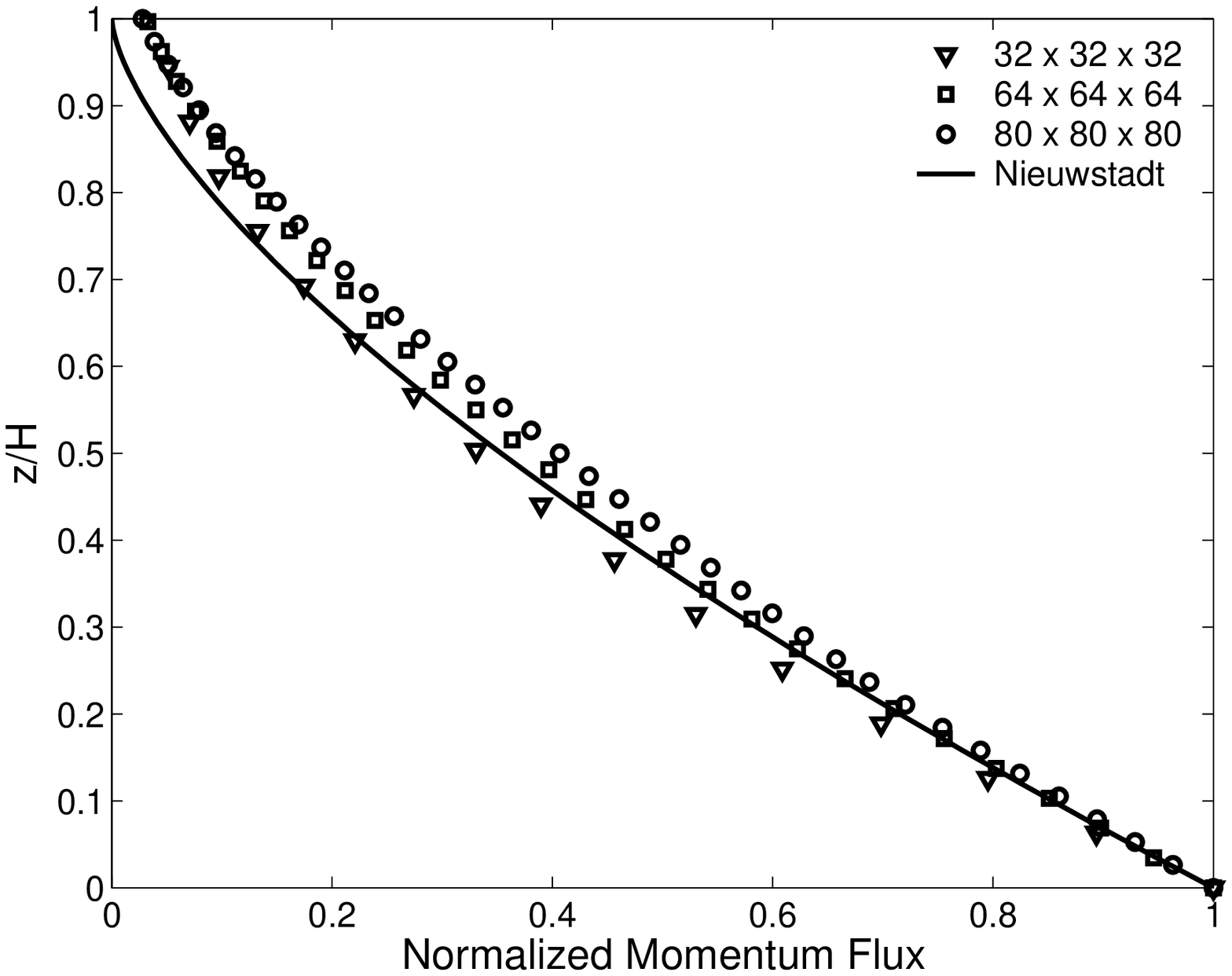}}
\figbox*{}{} {\epsfxsize=3.25 in \epsfbox{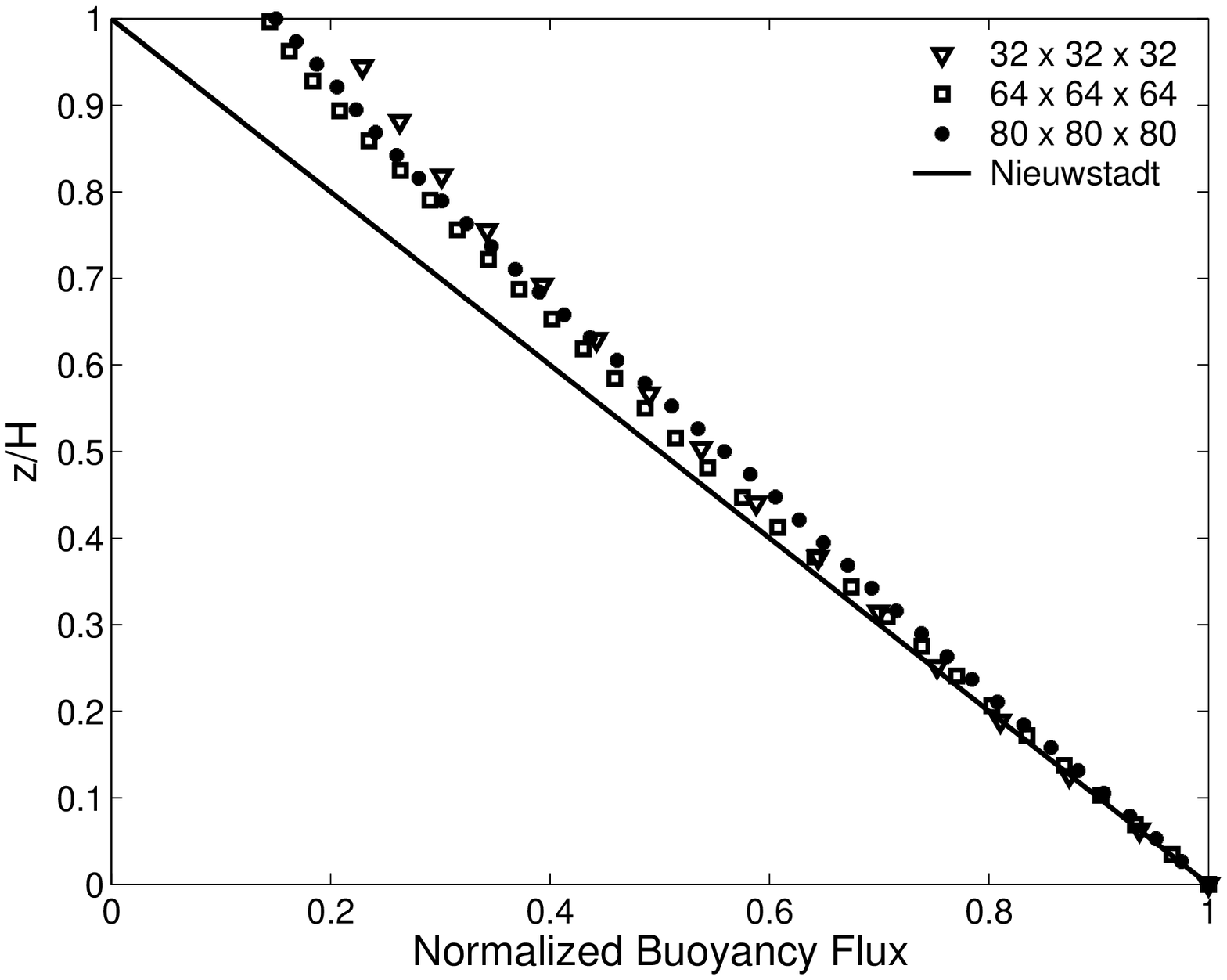}}
\caption{\label{FigStableNormFlux} Mean normalized momentum flux (top)
and normalized  buoyancy flux  (bottom) profiles from  the simulations
performed in the  present work.  These profiles are  averaged over the
last   one   hour   of   simulation.    Theoretical   predictions   by
\citeauthor{nieu85}    (\citeyear{nieu85})   are   also    shown   for
comparison.}
\end{figure}

Perhaps more  interesting is to  explore the normalized  flux profiles
shown in Figure \ref{FigStableNormFlux}. Nieuwstadt's analytical model
predictions are as follows \cite{nieu85}:
\begin{equation}
\frac{u_{*L}^2}{u_*^2} = \left( 1 - z/H \right)^{3/2}
\end{equation}
\begin{equation}
\frac{\left\langle{wb}_L\right\rangle}{\left\langle{wb_s}\right\rangle} 
= \left( 1 - z/H \right).
\end{equation}
Our model  simulated results are in close  agreement with Nieuwstadt's
predictions.

\begin{figure}[tb]
\figbox*{}{} {\epsfxsize=3.25 in  \epsfbox{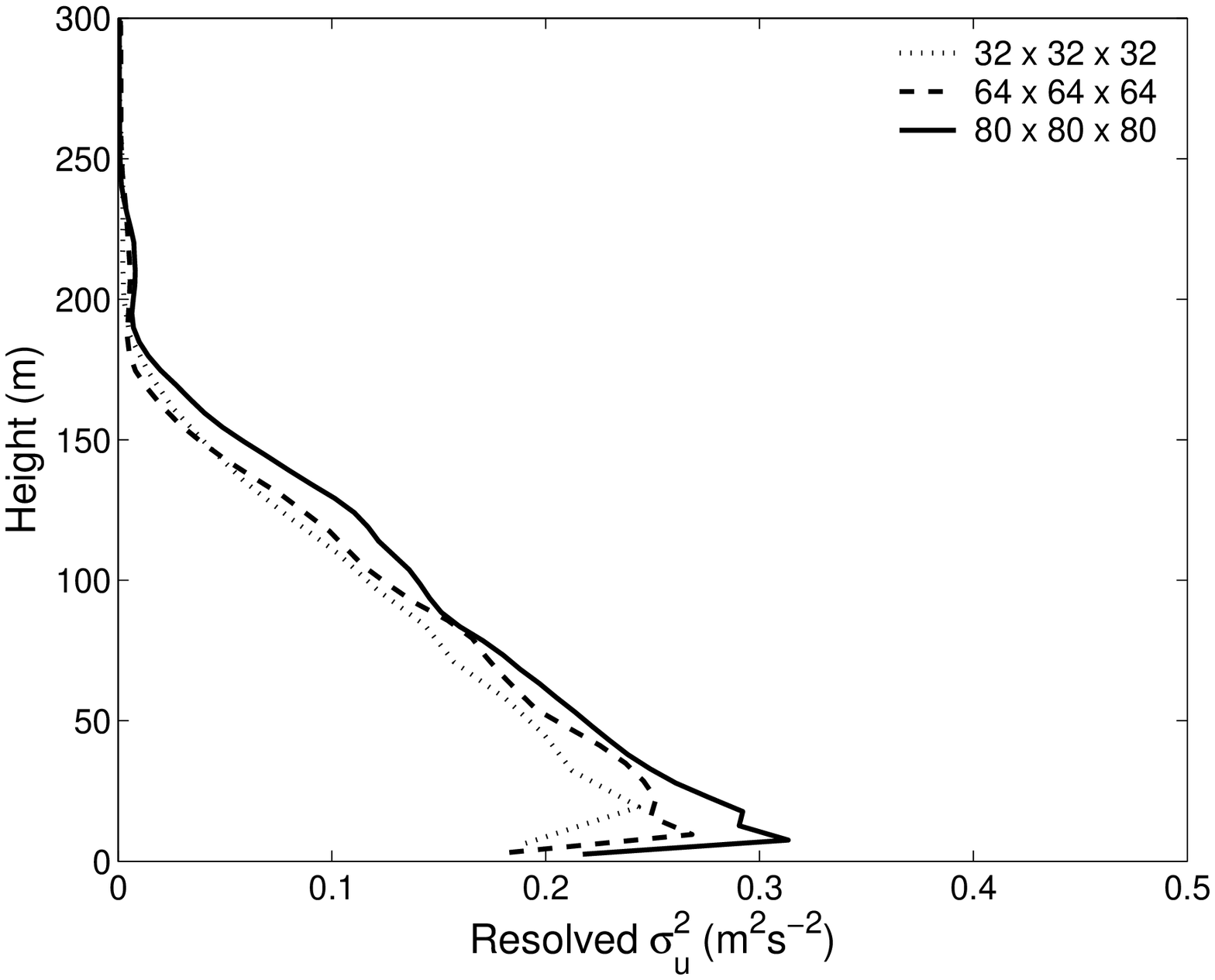}}
\figbox*{}{} {\epsfxsize=3.25 in  \epsfbox{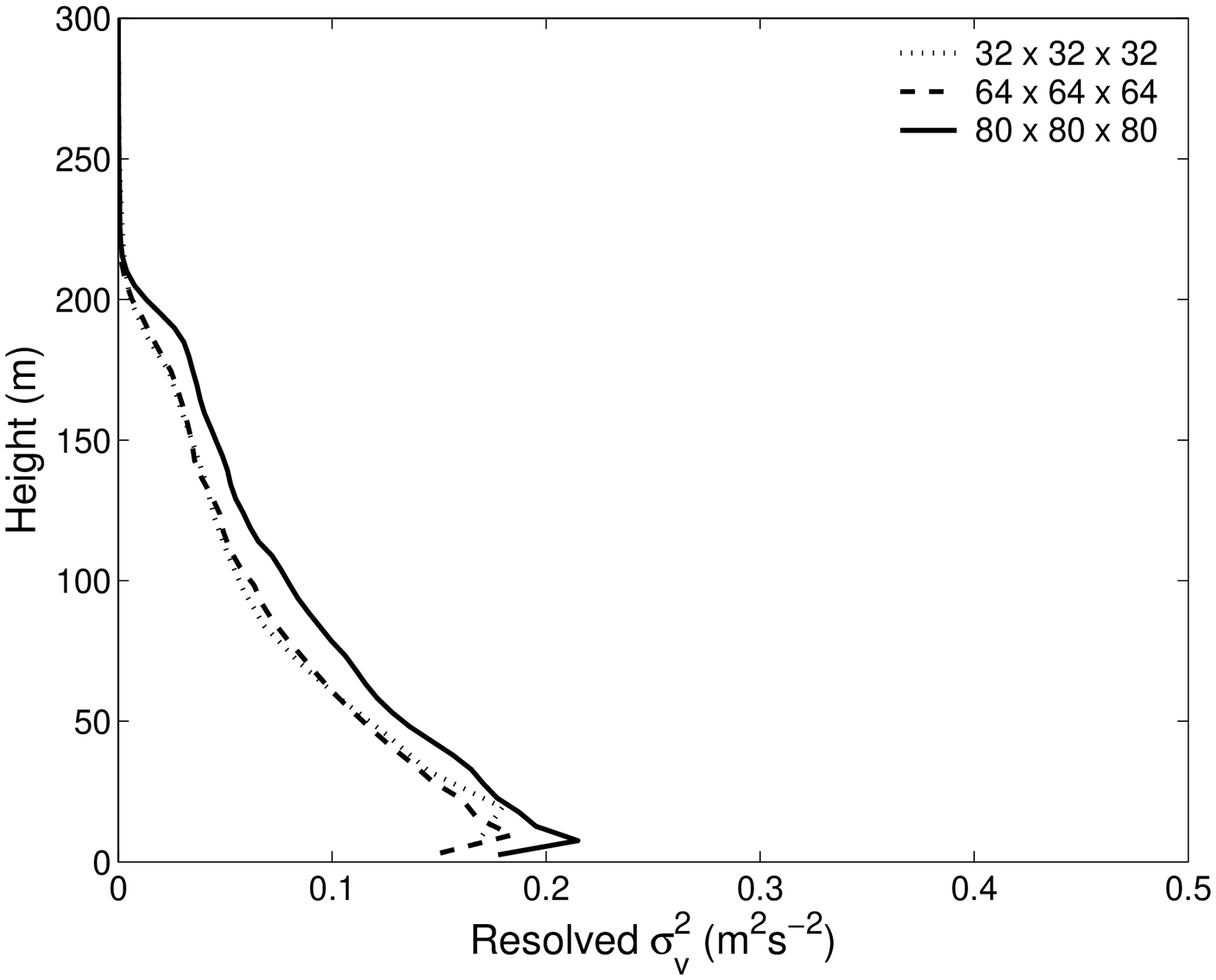}}
\figbox*{}{} {\epsfxsize=3.25 in  \epsfbox{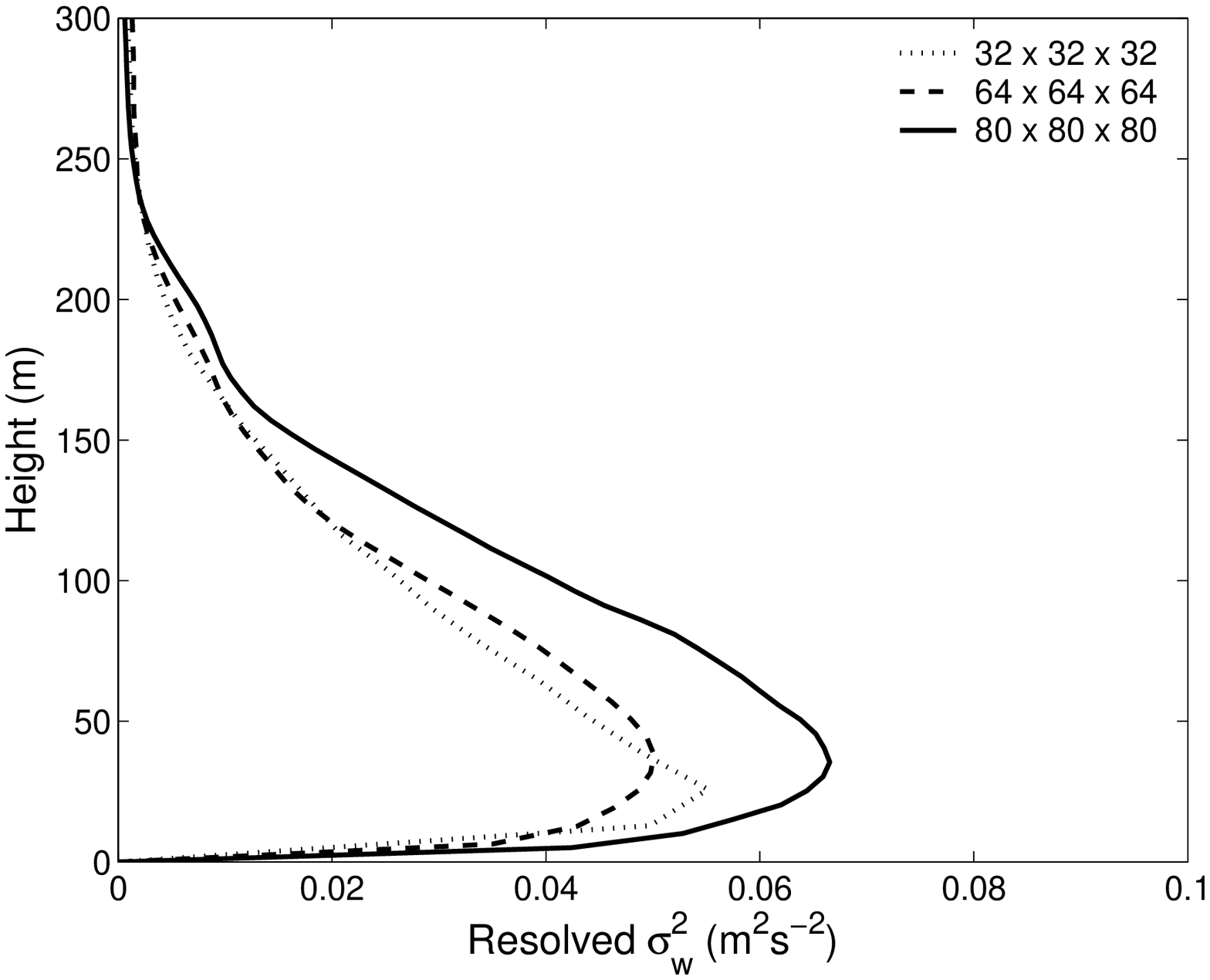}}
\caption{\label{FigStableVarU}  Resolved velocity  variances  from the
simulations performed in the present work. These profiles are averaged
over the last one hour of simulation.}
\end{figure}

\begin{figure}[tb]
\figbox*{}{}
{\epsfxsize=3.25 in  \epsfbox{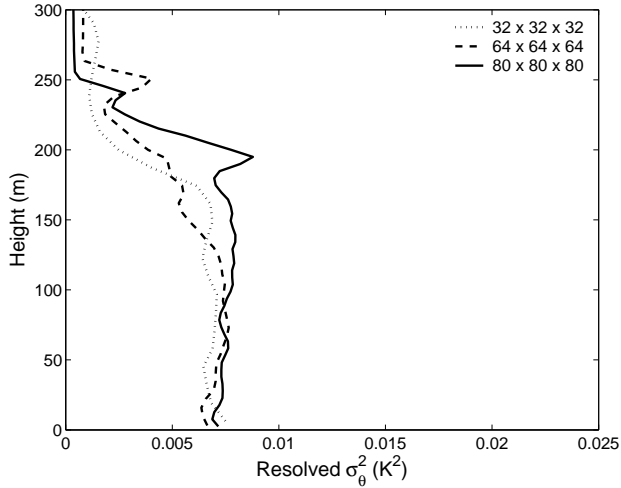}}
\caption{\label{FigStableVarT} Resolved temperature variances from the
simulations performed in the present work. These profiles are averaged
over the last one hour of simulation.}
\end{figure}

Figures   \ref{FigStableVarU}   and   \ref{FigStableVarT}   show   the
(resolved) variances  of velocity components and  temperature.  In the
surface layer, the normalized  resolved velocity variances are smaller
than Nieuwstadt's field  observations \cite{nieu84a,nieu84b}.  This is
expected as  the SGS  contributions to these  variances have  not been
added  here.  This  will  be done  while  studying Nieuwstadt's  local
scaling hypothesis and will be shown to be in excellent agreement with
Nieuwstadt's  observations (see below).   We would  like to  point out
that in  contrast to  our simulated results,  the nonlinear  SGS model
simulations  by  \citeauthor{koso00} (\citeyear{koso00})  surprisingly
yielded  surface layer  velocity variances  (resolved plus  SGS) which
were $\sim 40$ percent smaller than Nieuwstadt's observations.

In       his      local      scaling       hypothesis,      Nieuwstadt
\cite{nieu84a,nieu84b,nieu85}    conjectured    that   under    stable
stratification, the  local Obukhov  length ($\Lambda$) based  on local
turbulent  fluxes should be  considered as  a more  fundamental length
scale.  Then, according to this hypothesis, dimensionless combinations
of turbulent variables [gradients, fluxes, (co-)variances etc.]  which
are  measured  at  the  same   height  ($z$)  could  be  expressed  as
`universal'   functions    of   the   stability    parameter, $\zeta~
(=z/\Lambda)$. Exact  forms of these  functions could be  predicted by
dimensional analysis  only in the asymptotic very  stable case ($\zeta
\to \infty$),  as discussed below.  In the very stable  regime (z-less
condition),  since any  explicit dependence  on $z$  disappears, local
scaling    predicts    that    dimensionless   turbulent    quantities
asymptotically            approach           constant           values
\cite{nieu84a,nieu84b,nieu85}.  Local  scaling could  be  viewed as  a
generalization of the  well established Monin-Obukhov (M-O) similarity
theory \cite{moni71,sorb89}.  M-O similarity theory  is strictly valid
in the surface layer (lowest 10{\%} of the ABL), whereas local scaling
describes    the    turbulent   structure    of    the   entire    SBL
\cite{nieu84a,nieu84b,nieu85}.

Whether or not our  LES-generated statistics support the local scaling
hypothesis was  studied extensively in our  recent work \cite{basu05}.
In  that study,  we also  performed rigorous  statistical  analyses of
field observations and wind-tunnel measurements in order to verify the
validity of local scaling hypothesis under very stable conditions.  An
extensive  set  of  turbulence  statistics, computed  from  field  and
wind-tunnel  measurements   and  also  from   LES-generated  datasets,
supported the validity  of the local scaling hypothesis  (in the cases
of traditional bottom-up as well as upside-down stable boundary layers
over homogeneous,  flat terrains). We  demonstrated that non-turbulent
effects need to  be removed from field data  while studying similarity
hypotheses, otherwise  the results could  be misleading \cite{basu05}.
For completeness of the present  paper, we decided to include some key
local scaling results.  For  ease in representation, we categorize our
LES-generated database  based on local  stabilities ($z/\Lambda$) (see
Table \ref{LocSc_T2}). The class S1 represents near neutral stability;
while S5 corresponds to the very stable regime.

\begin{table}
\caption{Number of samples in each stability class.}
\label{LocSc_T2}
\begin{tabular}{lcccc} \hline\hline
Class & Stability ($\zeta$) & $32^3$ & $64^3$ & $80^3$ \\ 
\hline
S1 &  0.00-0.10 & 0 & 1 & 2 \\ 
S2 &  0.10-0.25 & 1 & 2 & 3 \\ 
S3 &  0.25-0.50 & 2 & 3 & 3 \\ 
S4 &  0.50-1.00 & 1 & 4 & 6 \\ 
S5 &  $>$~1.00  & 8 & 11 & 14 \\ 
\hline
\end{tabular} 
\end{table}

\begin{figure}[tb]
\figbox*{}{}  {\epsfxsize=3.25 in  \epsfbox{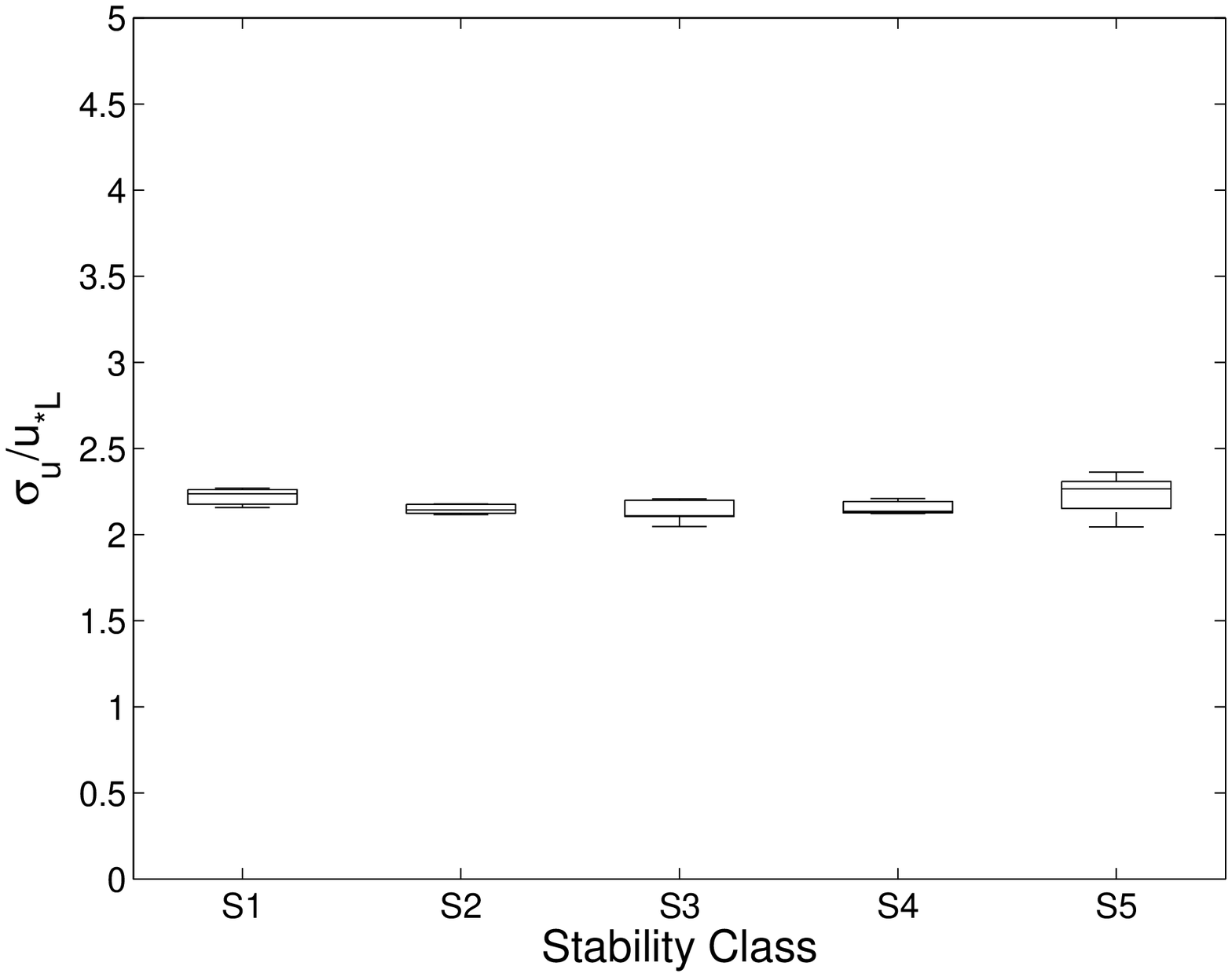}} \figbox*{}{}
{\epsfxsize=3.25      in      \epsfbox{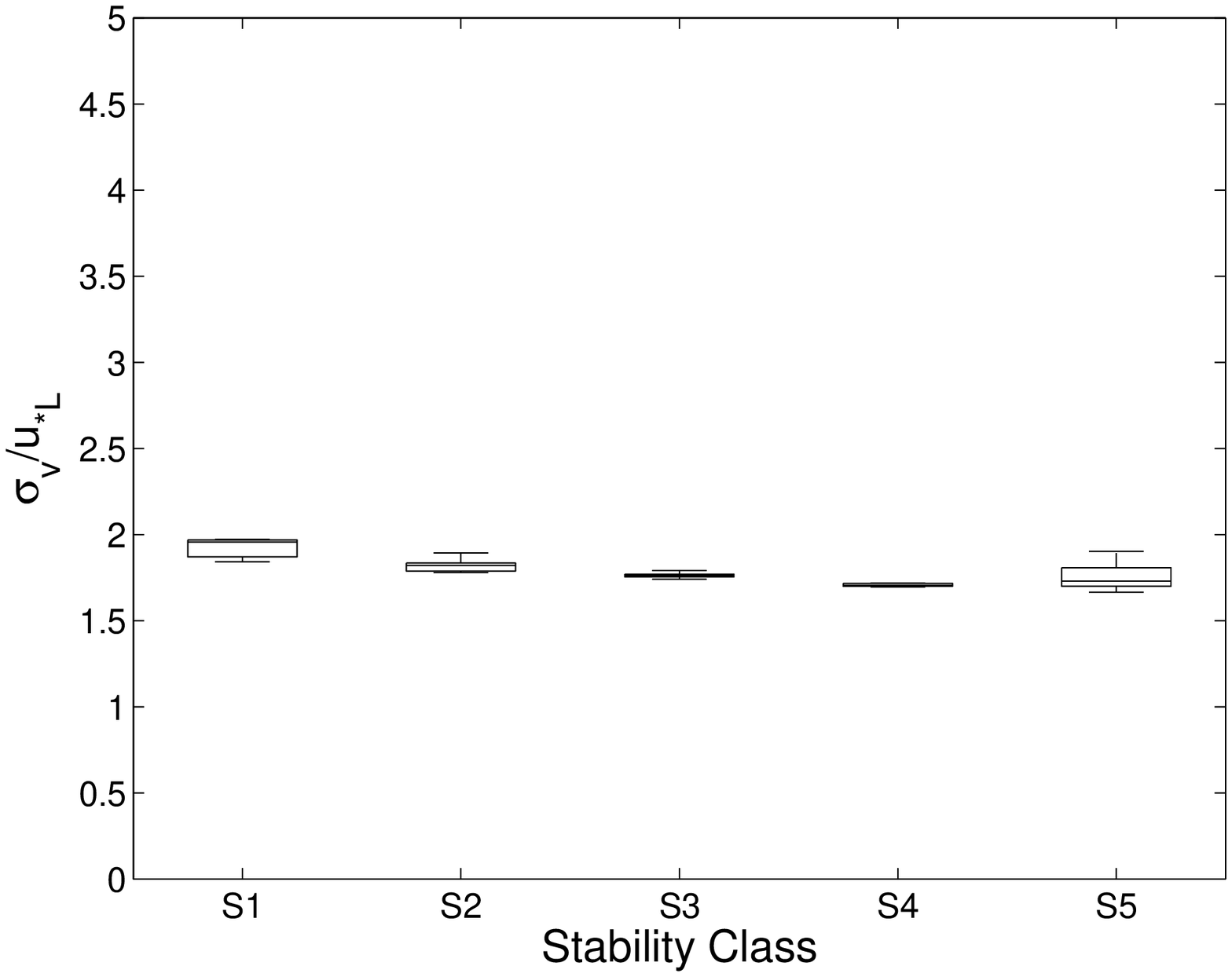}}      \figbox*{}{}
{\epsfxsize=3.25 in \epsfbox{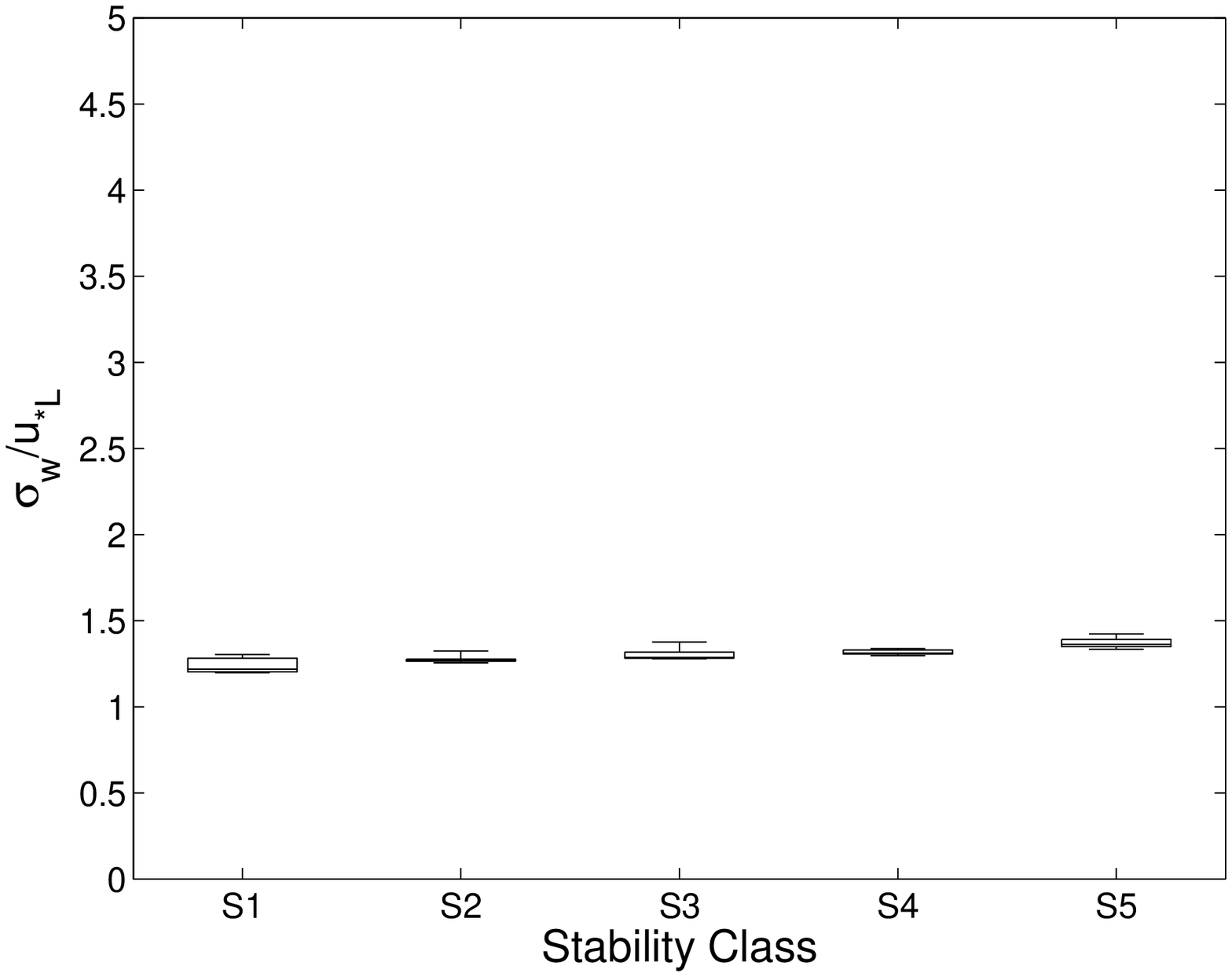}}
\caption{\label{LocSc_sigVel}     Normalized    longitudinal    (top),
transverse   (middle),  and   vertical   (bottom)  velocity   standard
deviations. These  statistics are averaged  over the last one  hour of
simulation.}
\end{figure}

\begin{figure}[tb]
\figbox*{}{} {\epsfxsize=3.25 in \epsfbox{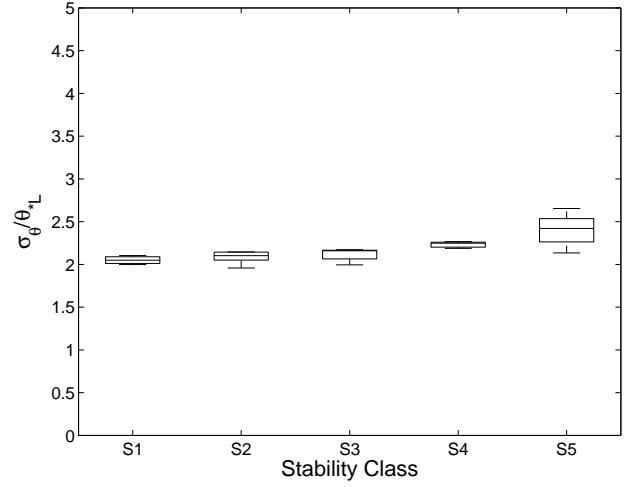}}
\caption{\label{LocSc_sigT} Normalized temperature standard deviations. 
This statistic is averaged over the last one hour of simulation.}
\end{figure}

In  Figures  \ref{LocSc_sigVel}   and  \ref{LocSc_sigT}  we  plot  the
normalized standard deviation  of velocity components and temperature,
respectively.   The  results  are  presented  using  standard  boxplot
notation  with  marks at  95,  75,  50, 25,  and  5  percentile of  an
empirical distribution.   The SGS contributions to  the total standard
deviations   are   estimated   following   the   approach   of   Mason
\cite{maso89,maso90}.

It   is   quite    evident   from   Figures   \ref{LocSc_sigVel}   and
\ref{LocSc_sigT}  that  the   normalized  standard  deviation  of  the
turbulence variables closely follows the local scaling predictions and
also z-less stratification. In  Table \ref{LocSc_T3} we further report
the median  values of the  turbulence statistics corresponding  to the
category S5. Loosely,  these median values could be  considered as the
asymptotic z-less  values, which are  found to be remarkably  close to
Nieuwstadt's analytical  predictions and also  field observations (see
Table \ref{LocSc_T3}).   For an example,  Nieuwstadt's theory predicts
that   the    normalized   vertical   velocity    standard   deviation
asymptotically  approaches   $\sim1.4$  in  the   z-less  regime.   In
\citeauthor{basu05}~(\citeyear{basu05}), we observed  this value to be
in  the narrow  range of  1.4 to  1.6.   Recently, \citeauthor{hein04}
(\citeyear{hein04}) compiled  a list (see  Table 2 of their  paper) of
turbulence statistics under very stable conditions ($\zeta_{max} \sim$
25) reported by  various researchers.  They found  an asymptotic value
of $\sim1.6$ for $\sigma_w/u_{*L}$, in accord with \citeauthor{sorb86}
(\citeyear{sorb86}).

\begin{figure}[tb]
\figbox*{}{}  {\epsfxsize=3.25  in \epsfbox{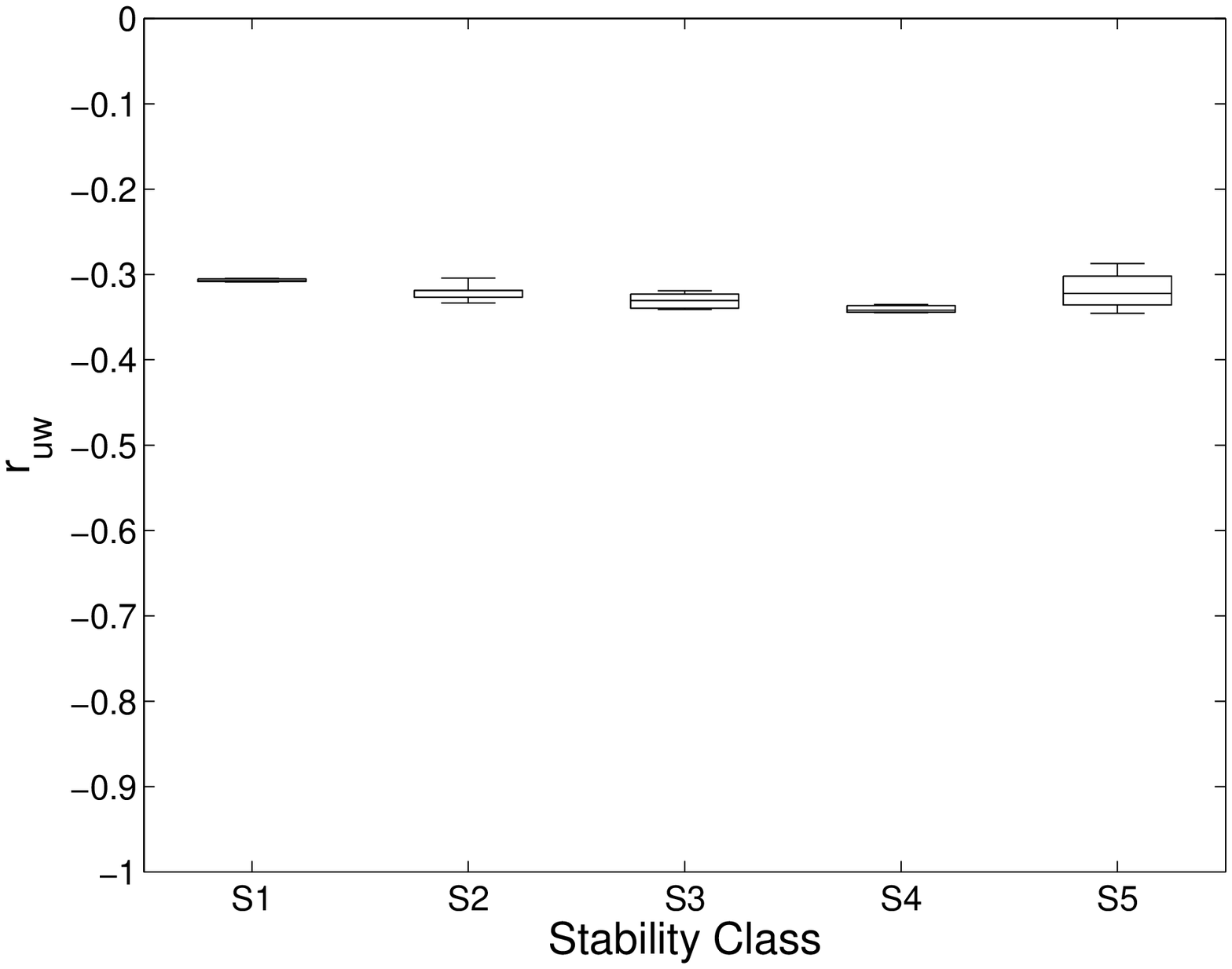}}  \figbox*{}{}
{\epsfxsize=3.25       in      \epsfbox{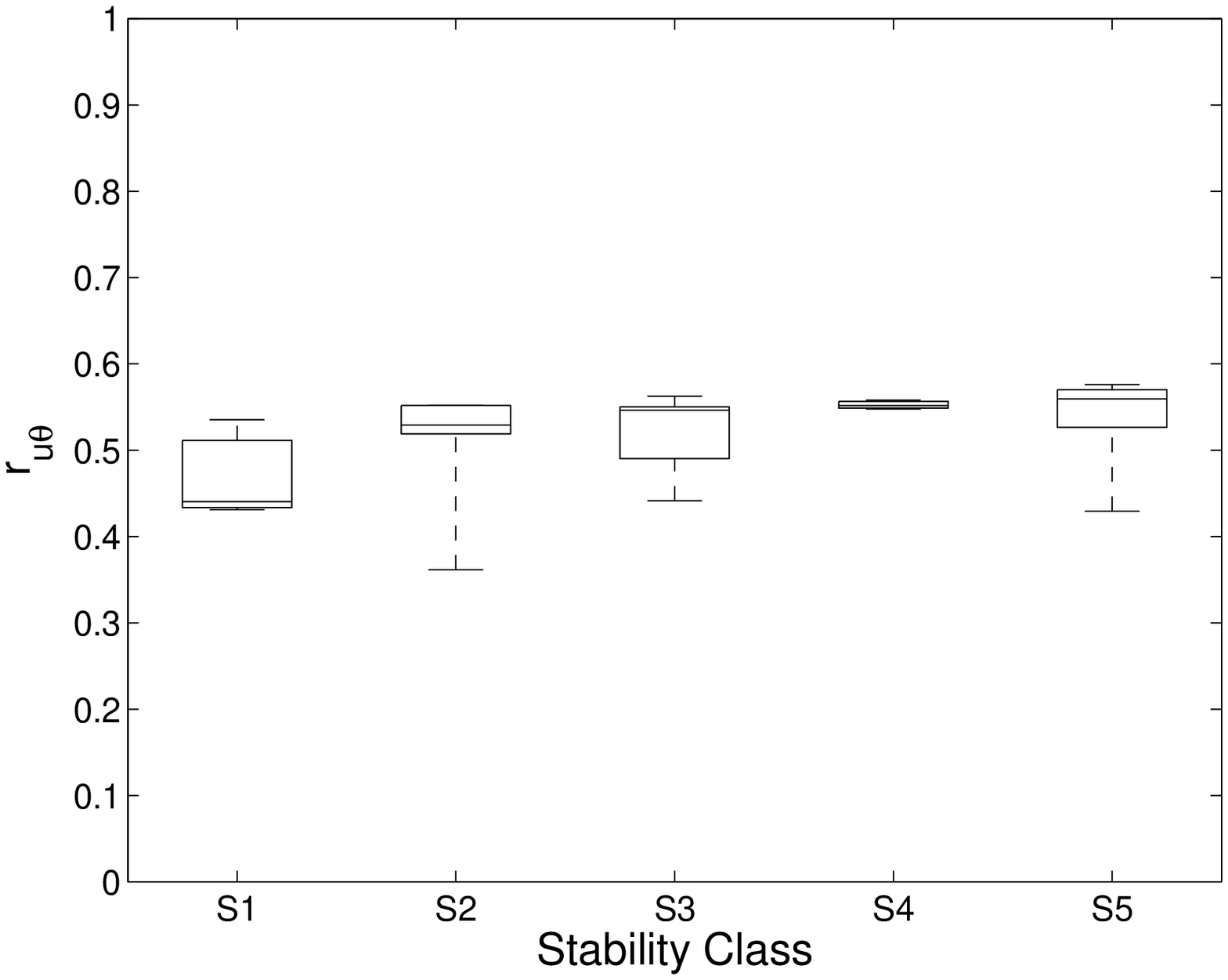}}      \figbox*{}{}
{\epsfxsize=3.25 in \epsfbox{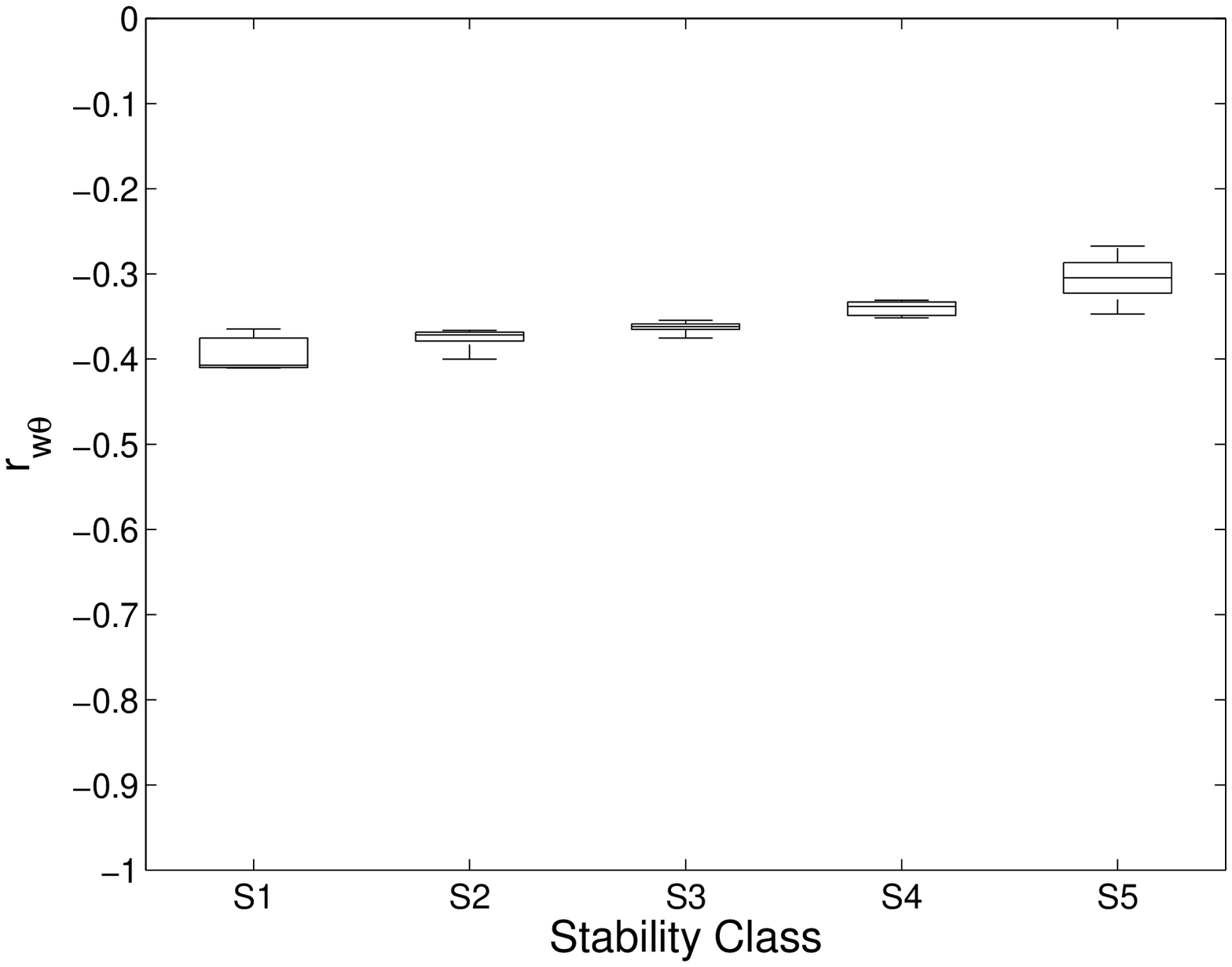}}
\caption{\label{LocSc_r}  Correlation between  $u$ and  $w$ ($r_{uw}$)
(top), $u$ and $\theta$ ($r_{u\theta}$) (middle), and $w$ and $\theta$
($r_{w\theta}$) (bottom). These statistics  are averaged over the last
one hour of simulation.}
\end{figure}

In  Figure \ref{LocSc_r},  we report  the mutual  correlations between
$u$, $w$ and  $\theta$.  The z-less values are  also reported in Table
\ref{LocSc_T3}. Once again, these values  are very similar to the ones
compiled  by  \citeauthor{hein04}  (\citeyear{hein04}),  our  previous
study              \cite{basu05},              results              of
\citeauthor{sorb86}~(\citeyear{sorb86}) and theoretical predictions of
\citeauthor{nieu84b}~(\citeyear{nieu84b}).        As      a      note,
\citeauthor{kaim94}~(\citeyear{kaim94})  also  report  that for  $0  <
\zeta < 1$, $r_{u\theta}$ = 0.6, which is close to the values found in
the present study.

\begin{table}
\caption{z-less values of turbulence statistics.}\label{LocSc_T3}
\begin{tabular}{lccccc} \hline \hline
Turbulence  &   Large-eddy  &  Nieuwstadt    & Sorbjan\\
Statistics  &   Simulations &  (1984b, 1985) & (1986)\\  
\hline 
$\sigma_u/u_{*L}$   &  2.3  &  2.0 & 2.4\\
$\sigma_v/u_{*L}$   &  1.7  &  1.7 & 1.8\\  
$\sigma_w/u_{*L}$   &  1.4  &  1.4 & 1.6\\ 
$\sigma_\theta/\theta_{*L}$ & 2.4 & 3.0 & 2.4\\ 
$r_{uw}$            & -0.32  & - & -\\ 
$r_{u\theta}$       & 0.56 & - & 0.50\\ 
$r_{w\theta}$       & -0.30 & -0.24 & -\\
\hline
\end{tabular} 
\end{table}

In  light of  the  foregoing analyses  it  is certain  that the  local
scaling  hypothesis of  Nieuwstadt, which  has survived  the  last two
decades,  still  holds  for  a  wide range  of  stabilities  and  is 
well reproduced by our LES model.

\subsubsection{SGS Coefficients}

\begin{figure}[tb]
\figbox*{}{} {\epsfxsize=3.25 in  \epsfbox{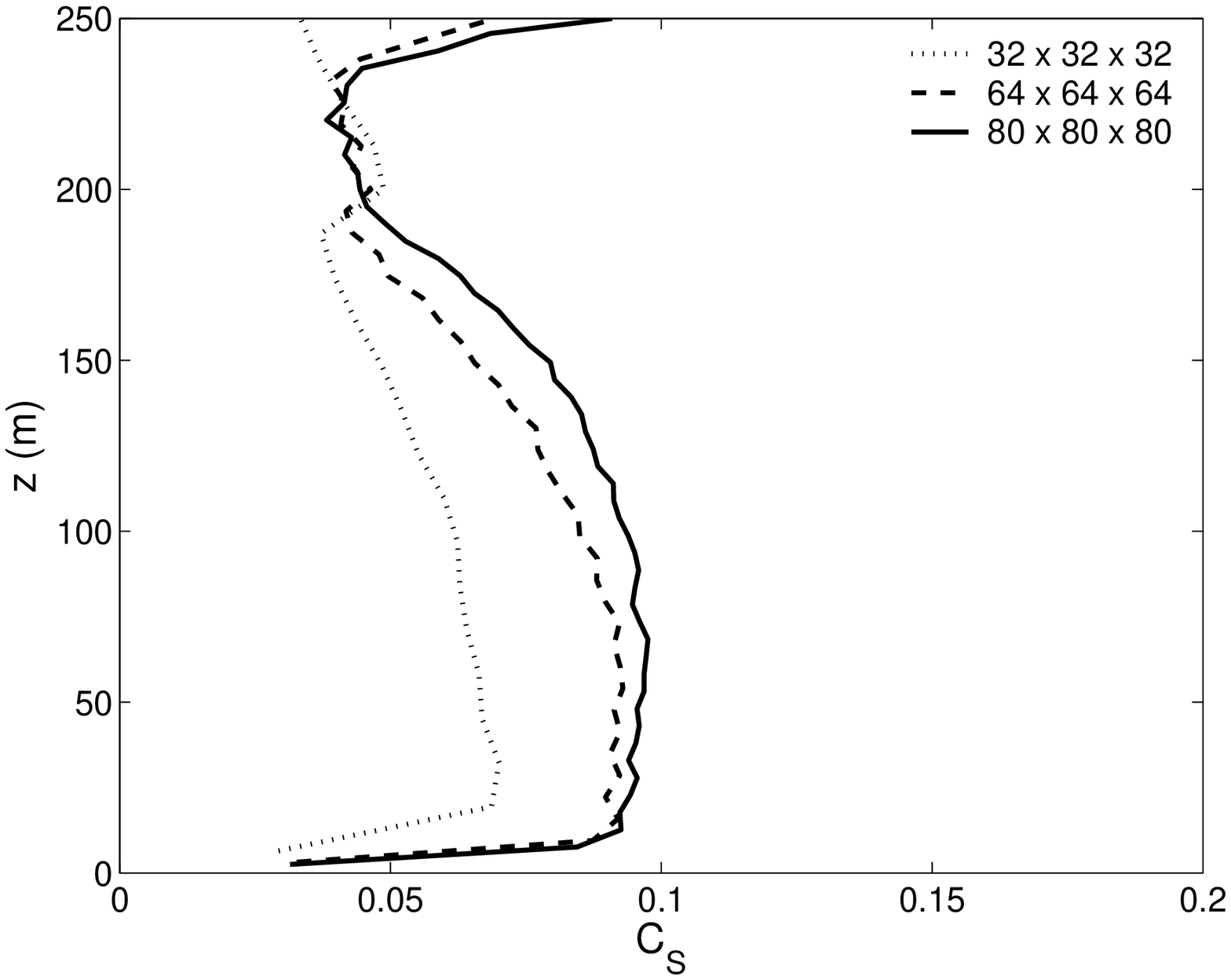}}
\figbox*{}{} {\epsfxsize=3.25 in  \epsfbox{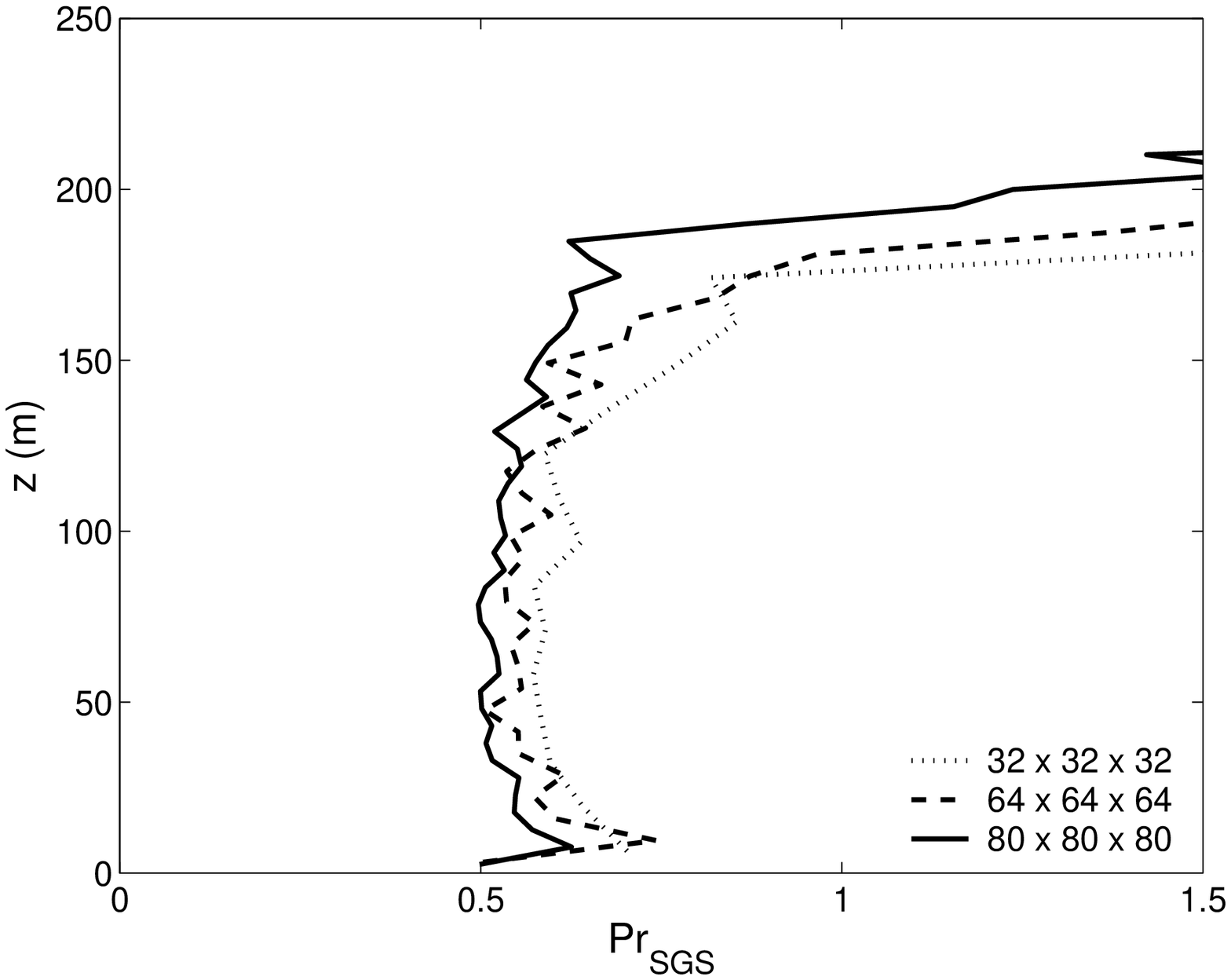}}
\caption{\label{FigStableSGS}    Vertical   profiles   of    the   SGS
coefficients:   $C_S$,  $Pr_{SGS}$,   dynamically   obtained  by   the
locally-averaged   scale-dependent  dynamic   (LASDD)   model.   These
profiles are averaged over the last one hour of simulation.}
\end{figure}

\begin{figure}[tb]
\figbox*{}{} {\epsfxsize=3.25 in  \epsfbox{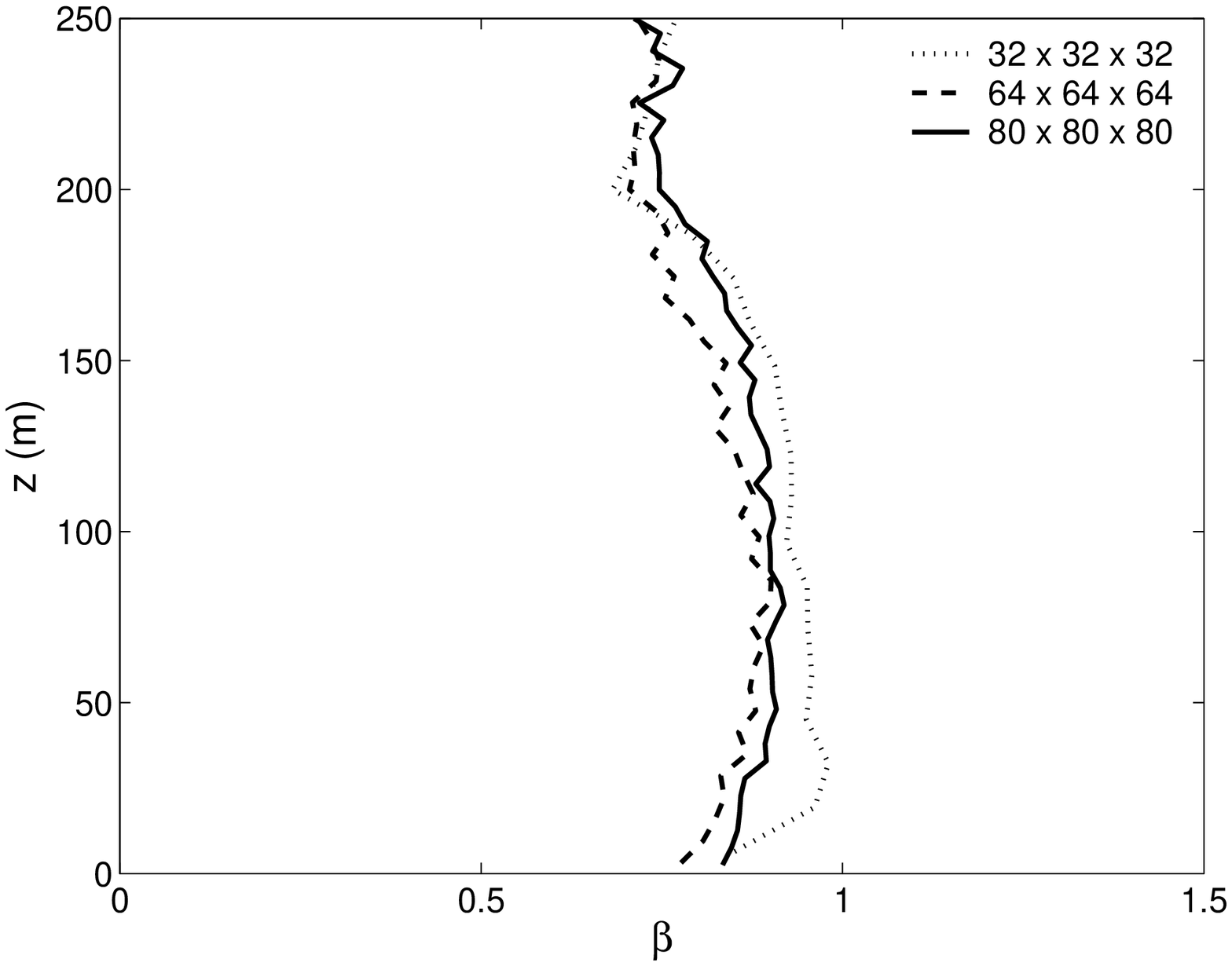}}
\figbox*{}{} {\epsfxsize=3.25 in  \epsfbox{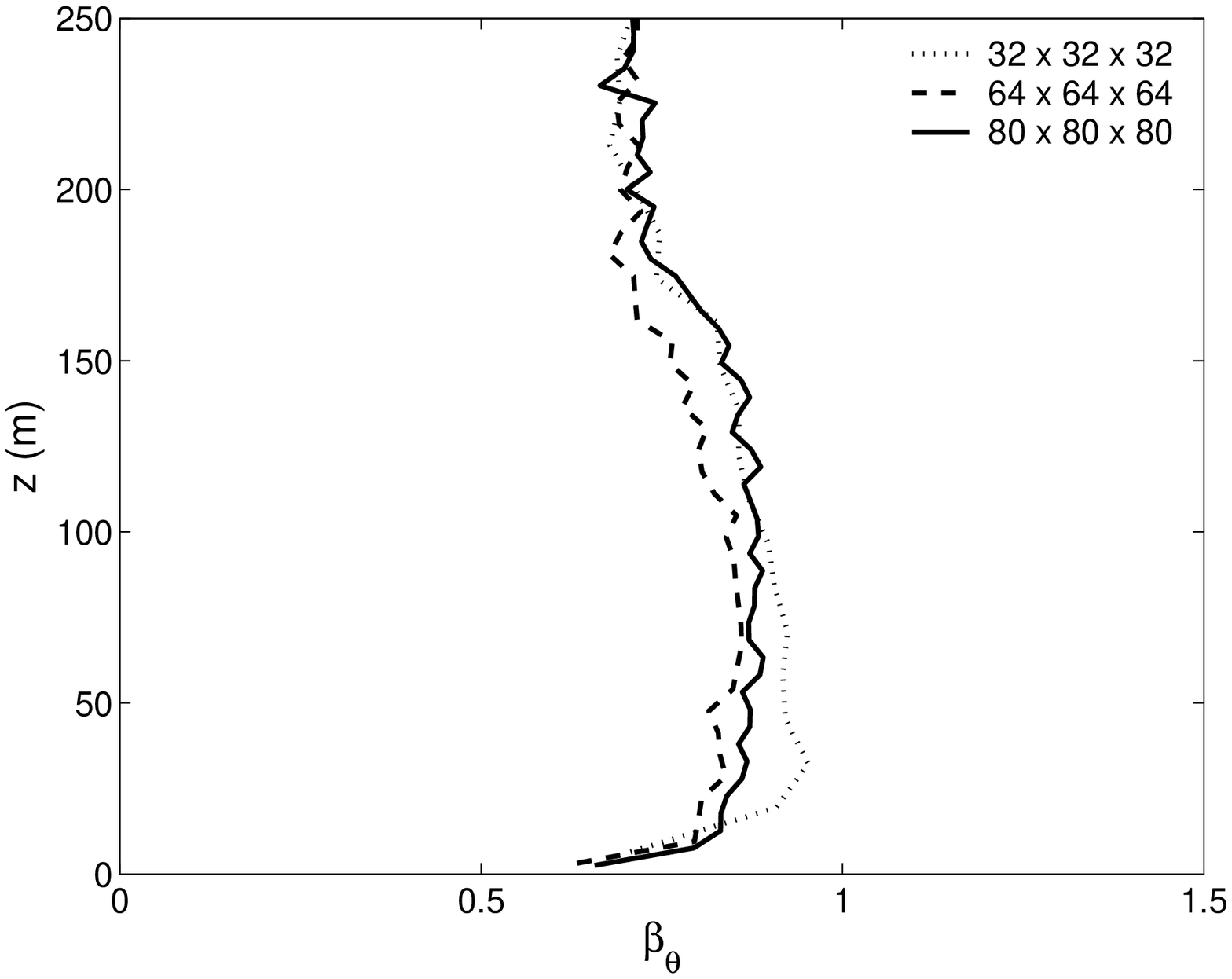}}
\caption{\label{FigStableBeta}     Vertical     profiles    of     the
scale-dependent   parameters:  $\beta$,   $\beta_\theta$,  dynamically
obtained  by  the  locally-averaged  scale-dependent  dynamic  (LASDD)
model.  These  profiles  are  averaged  over  the  last  one  hour  of
simulation.}
\end{figure}

Figures  \ref{FigStableSGS}  and   \ref{FigStableBeta}  show  the  SGS
coefficients:  $C_S$,  $Pr_{SGS}$,  and the  averaged  scale-dependent
parameters:  $\beta$, $\beta_\theta$,  dynamically obtained  using the
locally-averaged  scale-dependent dynamic model.  $\beta$
and $\beta_\theta$ are  found to be significantly smaller  than $1$ in
the entire boundary layer. This  stresses the fact that the assumption
of  scale-invariance in  anisotropic  stably stratified  flows is  not
appropriate. Indeed, both  $C_S$ and $Pr_{SGS}$ are found  to be scale
dependent. The scale-dependent parameters  are also expected to depend
on  local stability  as they  decrease significantly  in  the strongly
stratified  inversion   layer.   $C_S$  is  found   to  decrease  with
increasing  atmospheric   stability,  consistent  with   recent  field
observations          (\citeauthor{port01}          \citeyear{port01};
\citeauthor{klei03}   \citeyear{klei03}).   The  SGS   Prandtl  number
($Pr_{SGS}$) is  more or less  constant inside the boundary  layer and
gradually increases  to $\sim 1$  in the inversion layer,  as commonly
assumed.  Moreover,  the values of $Pr_{SGS}$ increase  in the surface
layer.  Earlier,  we described  very similar behavior  in the  case of
turbulent Prandtl number ($Pr_t$).

\section{Concluding Remarks and Future Perspectives}

One  of the  contributions of  this  research is  the development  and
implementation of a new-generation  subgrid-scale model, termed as the
locally-averaged  scale-dependent  dynamic  (LASDD) model.   This  SGS
model   shares  most   of   the  desirable   characteristics  of   the
plane-averaged scale-dependent dynamic  models, originally proposed by
\citeauthor{port00}~(\citeyear{port00})  for  the  SGS  stresses,  and
\citeauthor{port04}~(\citeyear{port04}) for  SGS fluxes.  For example,
it does  not require  any {\it a priori} specification of  SGS coefficients
since they are computed  dynamically in a self-consistent manner.  The
dynamically  estimated coefficients  are found  to strongly  depend on
filter scale and atmospheric stability, in close agreement with {\it a
priori}   field    studies   (\citeauthor{port01}   \citeyear{port01};
\citeauthor{klei03} \citeyear{klei03}).   However, in contrast  to the
original plane-averaged version, the  LASDD model does not suffer from
the  insufficient SGS  dissipation  problem in  simulations of  stable
boundary layers.

The  potential of  our SGS  model is  made clear  in coarse-resolution
large-eddy simulations of moderately stable boundary layers.  Overall,
the  agreements between  our LES-generated  turbulence  statistics and
observations, as well  as some well-established empirical formulations
and  theoretical predictions  are remarkable.   The results  also show
clear  improvements over  most of  the traditional  SGS models  in the
surface layer.  In essence,  we showed that tuning-free simulations of
stable atmospheric  boundary layers are feasible  even with relatively
coarse resolutions if one uses a robust and reliable SGS scheme.

The  next logical  step  would be  to  check the  performance of  this
new-generation SGS  scheme in simulating very  stable boundary layers.
This  would of  course require  extensive validation  against existing
profiles  of various turbulence  statistics measured  during different
field  campaigns (e.g,  Cooperative Atmosphere-Surface  Exchange Study
1999  -  CASES99; Beaufort  Sea  Arctic  Stratus  Experiment -  BASE;
Surface  Heat  Budget  of the  Arctic  Ocean  -  SHEBA).  One  of  the
characteristics  of   strongly  stratified  boundary   layers  is  the
existence of global intermittency (turbulent burstings in the midst of
a  laminar flow).   In contrast  to  the traditional  SGS models,  the
locally-averaged  scale-dependent dynamic  LES model  has  the correct
behavior in laminar and transitional flows. This makes us believe that
this model will be able to model the complex intermittency behavior of
the very stable boundary layer flows.

Contemporary SBL research has revealed that in very stable regimes the
relative importance  of radiative cooling  and heat exchange  with the
underlying   soil   becomes  as   significant   as  turbulent   mixing
\cite{vand02}.  This means that radiation and soil physics should also
be included  in LES models before attempting  very stable simulations.
Another interesting  feature of this  type of boundary layer  flows is
the  presence  of gravity  waves.   Conceptually,  LES  is capable  of
simulating gravity  waves, provided the  domain size is  large enough.
Unfortunately, the  present computational power dictates  that in such
cases one  must have  a relatively coarse  resolution making  thus the
locally-averaged scale-dependent dynamic model more desirable than the
laminarization-prone  traditional SGS  models.  These  issues  will be
addressed in our future research.

\begin{acknowledgements}
The authors are grateful to Rob Stoll for his generous help during the
course of this work.  We  thank Efi Foufoula-Georgiou for many thought
provoking discussions.  This work was partially funded by NSF and NASA
grants.  All  the computational resources were kindly  provided by the
National Center for Atmospheric Research (NCAR).
\end{acknowledgements}

\bibliography{SBL_JAS}

\begin{thebibliography}{74}
\expandafter\ifx\csname natexlab\endcsname\relax\def\natexlab#1{#1}\fi

\bibitem[{Albertson and Parlange(1999){\it Albertson and Parlange\/}}]{albe99}
Albertson, J.~D., and M.~B. Parlange, 1999:
\newblock Natural integration of scalar fluxes from complex terrain.
\newblock {\it Adv. Wat. Res.\/}, {\bf 23}, 239--252.

\bibitem[{Andr\'{e}n(1995){\it Andr\'{e}n\/}}]{andr95}
Andr\'{e}n, A., 1995:
\newblock The structure of stably stratified atmospheric boundary layers: A
  large-eddy simulation study.
\newblock {\it Quart. J. Roy. Meteorol. Soc.\/}, {\bf 121}, 961--985.

\bibitem[{Andr\'{e}n et~al.(1994){\it Andr\'{e}n, Brown, Graf, Mason, Moeng,
  Nieuwstadt, and Schumann\/}}]{andr94}
Andr\'{e}n, A., A.~R. Brown, J.~Graf, P.~J. Mason, C.-H. Moeng, F.~T.~M.
  Nieuwstadt, and U.~Schumann, 1994:
\newblock Large-eddy simulation of a neutrally stratified boundary layer: A
  comparison of four computer codes.
\newblock {\it Quart.~J.~Roy. Meteorol. Soc.\/}, {\bf 120}, 1457--1484.

\bibitem[{Arya(2001){\it Arya\/}}]{arya01}
Arya, S.~P., 2001:
\newblock {\it Introduction to Micrometeorology\/}.
\newblock Academic Press, 420 pp.

\bibitem[{Basu et~al.(2005){\it Basu, Port\'{e}-Agel, Foufoula-Georgiou,
  Vinuesa, and Pahlow\/}}]{basu05}
Basu, S., F.~Port\'{e}-Agel, E.~Foufoula-Georgiou, J.-F. Vinuesa, and
  M.~Pahlow, 2005:
\newblock Revisiting the local scaling hypothesis in stably stratified
  atmospheric boundary layer turbulence: an integration of field and laboratory
  measurements with large-eddy simulations.
\newblock {\it Boundary-Layer Meteorol\/}.
\newblock Under review.

\bibitem[{Beare and MacVean(2004){\it Beare and MacVean\/}}]{bear04a}
Beare, R.~J., and M.~K. MacVean, 2004:
\newblock Resolution sensitivity and scaling of large-eddy simulations of the
  stable boundary layer.
\newblock {\it Boundary-Layer Meteorol.\/}, {\bf 112}, 257--281.

\bibitem[{Beare et~al.(2004){\it Beare, et~al.\/}}]{bear04b}
Beare, R.~J., et~al., 2004:
\newblock An intercomparison of large-eddy simulations of the stable boundary
  layer.
\newblock {\it Boundary-Layer Meteorol\/}.
\newblock Under review.

\bibitem[{Beljaars and Holtslag(1991){\it Beljaars and Holtslag\/}}]{belj91}
Beljaars, A.~C.~M., and A.~A.~M. Holtslag, 1991:
\newblock Flux parameterization over land surfaces for atmospheric models.
\newblock {\it J. Appl. Meteorol.\/}, {\bf 30}, 327--341.

\bibitem[{Brost and Wyngaard(1978){\it Brost and Wyngaard\/}}]{bros78}
Brost, R.~A., and J.~C. Wyngaard, 1978:
\newblock A model study of the stably stratified planetary boundary layer.
\newblock {\it J. Atmos. Sci.\/}, {\bf 35}, 1427--1440.

\bibitem[{Brown et~al.(1994){\it Brown, Derbyshire, and Mason\/}}]{brow94}
Brown, A.~R., S.~H. Derbyshire, and P.~J. Mason, 1994:
\newblock Large-eddy simulation of stable atmospheric boundary layers with a
  revised stochastic subgrid model.
\newblock {\it Quart. J. Roy. Meteorol. Soc.\/}, {\bf 120}, 1485--1512.

\bibitem[{Businger et~al.(1971){\it Businger, Wyngaard, Izumi, and
  Bradley\/}}]{busi71}
Businger, J.~A., J.~C. Wyngaard, Y.~Izumi, and E.~F. Bradley, 1971:
\newblock Flux-profile relationships in the atmospheric surface layer.
\newblock {\it J. Atmos. Sci.\/}, {\bf 28}, 181--189.

\bibitem[{Canuto et~al.(1988){\it Canuto, Hussaini, Quarteroni, and
  Zhang\/}}]{canu88}
Canuto, C., M.~Y. Hussaini, A.~Quarteroni, and T.~A. Zhang, 1988:
\newblock {\it Spectral Methods in Fluid Dynamics\/}.
\newblock Springer Verlag, 557 pp.

\bibitem[{Canuto and Cheng(1997){\it Canuto and Cheng\/}}]{canu97}
Canuto, V.~M., and Y.~Cheng, 1997:
\newblock Determination of the {Smagorinsky-Lilly} constant {$C_S$}.
\newblock {\it Phys. Fluids\/}, {\bf 9}, 1368--1378.

\bibitem[{Deardorff(1970){\it Deardorff\/}}]{dear70b}
Deardorff, J.~W., 1970:
\newblock Convective velocity and temperature scales for the unstable planetary
  boundary layer and {Rayleigh} convection.
\newblock {\it J. Atmos. Sci.\/}, {\bf 27}, 1211--1213.

\bibitem[{Deardorff(1972){\it Deardorff\/}}]{dear72}
Deardorff, J.~W., 1972:
\newblock Theoretical expression for the countergradient vertical heat flux.
\newblock {\it J. Geophys. Res.\/}, {\bf 77}, 5900--5904.

\bibitem[{Deardorff(1974){\it Deardorff\/}}]{dear74}
Deardorff, J.~W., 1974:
\newblock Three-dimensional numerical study of turbulence in an entraining
  mixed layer.
\newblock {\it Boundary-Layer Meteorol.\/}, {\bf 7}, 199--226.

\bibitem[{Deardorff(1980){\it Deardorff\/}}]{dear80}
Deardorff, J.~W., 1980:
\newblock Stratocumulus-capped mixed layers derived from a three-dimensional
  model.
\newblock {\it Boundary-Layer Meteorol.\/}, {\bf 18}, 495--527.

\bibitem[{Derbyshire(1990){\it Derbyshire\/}}]{derb90}
Derbyshire, S.~H., 1990:
\newblock Nieuwstadt's stable boundary layer revisited.
\newblock {\it Quart. J. Roy. Meteorol. Soc.\/}, {\bf 116}, 127--158.

\bibitem[{Derbyshire(1999){\it Derbyshire\/}}]{derb99}
Derbyshire, S.~H., 1999:
\newblock Stable boundary-layer modelling: Established approaches and beyond.
\newblock {\it Boundary-Layer Meteorol.\/}, {\bf 90}, 423--446.

\bibitem[{Germano(1992){\it Germano\/}}]{germ92}
Germano, M., 1992:
\newblock Turbulence: the filtering approach.
\newblock {\it J. Fluid Mech.\/}, {\bf 238}, 325--336.

\bibitem[{Germano et~al.(1991){\it Germano, Piomelli, Moin, and
  Cabot\/}}]{germ91}
Germano, M., U.~Piomelli, P.~Moin, and W.~H. Cabot, 1991:
\newblock A dynamic subgrid-scale eddy viscosity model.
\newblock {\it Phys. Fluids A\/}, {\bf 3}, 1760--1765.

\bibitem[{Geurts(2003){\it Geurts\/}}]{geur03}
Geurts, B.~J., 2003:
\newblock {\it Elements of Direct and Large-eddy Simulation\/}.
\newblock Edwards, 329 pp.

\bibitem[{Ghosal et~al.(1995){\it Ghosal, Lund, Moin, and
  Akselvoll\/}}]{ghos95}
Ghosal, S., T.~S. Lund, P.~Moin, and K.~Akselvoll, 1995:
\newblock A dynamic localization model for large-eddy simulation of turbulent
  flows.
\newblock {\it Phys. Fluids A\/}, {\bf 3}, 1760--1765.

\bibitem[{Heinemann(2004){\it Heinemann\/}}]{hein04}
Heinemann, G., 2004:
\newblock Local similarity properties of the continuously turbulent stable
  boundary layer over {Greenland}.
\newblock {\it Boundary-Layer Meteorol.\/}, {\bf 112}, 283--305.

\bibitem[{Higgins et~al.(2003){\it Higgins, Parlange, and Meneveau\/}}]{higg03}
Higgins, C., M.~B. Parlange, and C.~Meneveau, 2003:
\newblock Alignment trends of velocity gradients and subgrid scale fluxes in
  the turbulent atmospheric boundary layer.
\newblock {\it Boundary-Layer Meteorol.\/}, {\bf 109}, 59--83.

\bibitem[{Holtslag(2003){\it Holtslag\/}}]{holt03}
Holtslag, A. A.~M., 2003:
\newblock {GABLS} initiates intercomparison for stable boundary layer case.
\newblock {\it GEWEX News\/}, {\bf 13}, 7--8.

\bibitem[{Howell and Sun(1999){\it Howell and Sun\/}}]{howe99}
Howell, J.~F., and J.~Sun, 1999:
\newblock Surface-layer fluxes in stable conditions.
\newblock {\it Boundary-Layer Meteorol.\/}, {\bf 90}, 495--520.

\bibitem[{Hunt et~al.(1988){\it Hunt, Stretch, and Britter\/}}]{hunt88}
Hunt, J.~C.~R., D.~D. Stretch, and R.~E. Britter, 1988:
\newblock Length scales in stably stratified turbulent flows and their use in
  turbulence models.
\newblock  {\it Stably Stratified Flows and Dense Gas Dispersion\/}, J.~S.
  Puttock, Ed., Clarendon Press, pp. 285--321.

\bibitem[{Hunt et~al.(1996){\it Hunt, Shutts, and Derbyshire\/}}]{hunt96}
Hunt, J. C.~R., G.~J. Shutts, and S.~H. Derbyshire, 1996:
\newblock Stably stratified flows in meteorology.
\newblock {\it Dyn. Atmos. and Oceans\/}, {\bf 23}, 63--79.

\bibitem[{Kaimal and Finnigan(1994){\it Kaimal and Finnigan\/}}]{kaim94}
Kaimal, J.~C., and J.~J. Finnigan, 1994:
\newblock {\it Atmospheric Boundary Layer Flows: Their Structure and
  Measurement\/}.
\newblock Oxford University Press, 289 pp.

\bibitem[{Kleissl et~al.(2003){\it Kleissl, Meneveau, and Parlange\/}}]{klei03}
Kleissl, J., C.~Meneveau, and M.~B. Parlange, 2003:
\newblock On the magnitude and variability of subgrid-scale eddy-diffusion
  coefficients in the atmospheric surface layer.
\newblock {\it J. Atmos. Sci.\/}, {\bf 60}, 2372--2388.

\bibitem[{Kleissl et~al.(2004){\it Kleissl, Parlange, and
  Meneveau\/}}]{klei04a}
Kleissl, J., M.~B. Parlange, and C.~Meneveau, 2004:
\newblock Field experimental study of dynamic {Smagorinsky} models in the
  atmospheric surface layer.
\newblock {\it J. Atmos. Sci.\/}, {\bf 61}, 2296--2307.

\bibitem[{Kosovi\'{c}(1997){\it Kosovi\'{c}\/}}]{koso97}
Kosovi\'{c}, B., 1997:
\newblock Subgrid-scale modelling for the large-eddy simulation of
  high-{Reynolds}-number boundary layers.
\newblock {\it J. Fluid Mech.\/}, {\bf 338}, 151--182.

\bibitem[{Kosovi\'{c} and Curry(2000){\it Kosovi\'{c} and Curry\/}}]{koso00}
Kosovi\'{c}, B., and J.~A. Curry, 2000:
\newblock A large eddy simulation study of a quasi-steady, stably stratified
  atmospheric boundary layer.
\newblock {\it J. Atmos. Sci.\/}, {\bf 57}, 1052--1068.

\bibitem[{Lilly(1967){\it Lilly\/}}]{lill67}
Lilly, D.~K., 1967:
\newblock The representation of small-scale turbulence in numerical simulation
  experiments.
\newblock  {\it Proc. IBM Scientific Computing Symposium on Environmental
  Sciences\/}, pp. 195--210.

\bibitem[{Lilly(1992){\it Lilly\/}}]{lill92}
Lilly, D.~K., 1992:
\newblock A proposed modification of the {Germano} subgrid-scale closure
  method.
\newblock {\it Phys. Fluids A\/}, {\bf 4}, 633--635.

\bibitem[{Mahrt(1998){\it Mahrt\/}}]{mahr98a}
Mahrt, L., 1998:
\newblock Stratified atmospheric boundary layers and breakdown of models.
\newblock {\it Theoret. Comput. Fluid Dyn.\/}, {\bf 11}, 263--279.

\bibitem[{Mahrt and Vickers(2003){\it Mahrt and Vickers\/}}]{mahr03}
Mahrt, L., and D.~Vickers, 2003:
\newblock Formulation of turbulent fluxes in the stable boundary layer.
\newblock {\it J. Atmos. Sci.\/}, {\bf 60}, 2538--2548.

\bibitem[{Mason(1989){\it Mason\/}}]{maso89}
Mason, P., 1989:
\newblock Large-eddy simulation of the convective atmospheric boundary layer.
\newblock {\it J. Atmos. Sci.\/}, {\bf 46}, 1492--1516.

\bibitem[{Mason(1994){\it Mason\/}}]{maso94}
Mason, P., 1994:
\newblock Large-eddy simulation: A critical review of the technique.
\newblock {\it Quart. J. Roy. Meteorol. Soc.\/}, {\bf 120}, 1--26.

\bibitem[{Mason and Brown(1999){\it Mason and Brown\/}}]{maso99}
Mason, P.~J., and A.~R. Brown, 1999:
\newblock On subgrid models and filter operations in large-eddy simulations.
\newblock {\it J. Atmos. Sci.\/}, {\bf 56}, 2101--2114.

\bibitem[{Mason and Derbyshire(1990){\it Mason and Derbyshire\/}}]{maso90}
Mason, P.~J., and S.~H. Derbyshire, 1990:
\newblock Large-eddy simulation of the stably-stratified atmospheric boundary
  layer.
\newblock {\it Boundary-Layer Meteorol.\/}, {\bf 53}, 117--162.

\bibitem[{Meneveau and Katz(2000){\it Meneveau and Katz\/}}]{mene00}
Meneveau, C., and J.~Katz, 2000:
\newblock Scale-invariance and turbulence models for large-eddy simulation.
\newblock {\it Ann. Rev. Fluid Mech.\/}, {\bf 32}, 1--32.

\bibitem[{Meneveau et~al.(1996){\it Meneveau, Lund, and Cabot\/}}]{mene96}
Meneveau, C., T.~S. Lund, and W.~H. Cabot, 1996:
\newblock A {Lagrangian} dynamic subgrid-scale model of turbulence.
\newblock {\it J. Fluid Mech.\/}, {\bf 319}, 353--385.

\bibitem[{M\'{e}tais and Lesieur(1992){\it M\'{e}tais and Lesieur\/}}]{meta92}
M\'{e}tais, O., and M.~Lesieur, 1992:
\newblock Spectral large-eddy simulations of isotropic and stably-stratified
  turbulence.
\newblock {\it J. Fluid Mech.\/}, {\bf 239}, 157--194.

\bibitem[{Moeng(1984){\it Moeng\/}}]{moen84}
Moeng, C.-H., 1984:
\newblock A large-eddy simulation model for the study of planetary
  boundary-layer turbulence.
\newblock {\it J. Atmos. Sci.\/}, {\bf 41}, 2052--2062.

\bibitem[{Moin et~al.(1991){\it Moin, Squires, Cabot, and Lee\/}}]{moin91}
Moin, P., K.~Squires, W.~Cabot, and S.~Lee, 1991:
\newblock A dynamic subgrid-scale model for compressible turbulence and scalar
  transport.
\newblock {\it Phys. Fluids A\/}, {\bf 3}, 2746--2757.

\bibitem[{Monin and Yaglom(1971){\it Monin and Yaglom\/}}]{moni71}
Monin, A.~S., and A.~M. Yaglom, 1971:
\newblock {\it Statistical Fluid Mechanics: Mechanics of Turbulence\/}.
\newblock vol.~1, MIT Press, 769 pp.

\bibitem[{Nieuwstadt(1984a){\it Nieuwstadt\/}}]{nieu84a}
Nieuwstadt, F. T.~M., 1984a:
\newblock Some aspects of the turbulent stable boundary layer.
\newblock {\it Boundary-Layer Meteorol.\/}, {\bf 30}, 31--55.

\bibitem[{Nieuwstadt(1984b){\it Nieuwstadt\/}}]{nieu84b}
Nieuwstadt, F. T.~M., 1984b:
\newblock The turbulent structure of the stable, nocturnal boundary layer.
\newblock {\it J. Atmos. Sci.\/}, {\bf 41}, 2202--2216.

\bibitem[{Nieuwstadt(1985){\it Nieuwstadt\/}}]{nieu85}
Nieuwstadt, F. T.~M., 1985:
\newblock A model for the stationary, stable boundary layer.
\newblock  {\it Turbulence and Diffusion in Stable Environments\/}, J.~C.~R.
  Hunt, Ed., Clarendon Press, pp. 149--179.

\bibitem[{Nieuwstadt et~al.(1991){\it Nieuwstadt, Mason, Moeng, and
  Schumann\/}}]{nieu91}
Nieuwstadt, F.~T.~M., P.~J. Mason, C.-H. Moeng, and U.~Schumann, 1991:
\newblock Large-eddy simulation of the convective boundary layer: a comparison
  of four computer codes.
\newblock  {\it Turbulent Shear Flows 8\/}, F.~Durst, R.~Friedrich, B.~E.
  Launder, F.~W. Schmidt, U.~Schumann, and J.~H. Whitelaw, Eds., Springer, pp.
  343--367.

\bibitem[{Orszag and Pao(1974){\it Orszag and Pao\/}}]{orsz74}
Orszag, S.~A., and Y.-H. Pao, 1974:
\newblock Numerical computation of turbulent shear flows.
\newblock {\it Adv. Geophys.\/}, {\bf 18A}, 224--236.

\bibitem[{Piomelli and Liu(1995){\it Piomelli and Liu\/}}]{piom95}
Piomelli, U., and J.~Liu, 1995:
\newblock Large-eddy simulation of rotating channel flows using a localized
  dynamic model.
\newblock {\it Phys. Fluids\/}, {\bf 7}, 839--848.

\bibitem[{Port\'{e}-Agel(2004){\it Port\'{e}-Agel\/}}]{port04}
Port\'{e}-Agel, F., 2004:
\newblock A scale-dependent dynamic model for scalar transport in {LES} of the
  atmospheric boundary layer.
\newblock {\it Boundary-Layer Meteorol.\/}, {\bf 112}, 81--105.

\bibitem[{Port{\'{e}}-Agel et~al.(2000){\it Port{\'{e}}-Agel, Meneveau, and
  Parlange\/}}]{port00}
Port{\'{e}}-Agel, F., C.~Meneveau, and M.~B. Parlange, 2000:
\newblock A scale-dependent dynamic model for large-eddy simulations:
  Application to a neutral atmospheric boundary layer.
\newblock {\it J. Fluid Mech.\/}, {\bf 415}, 261--284.

\bibitem[{Port\'{e}-Agel et~al.(2001){\it Port\'{e}-Agel, Parlange, Meneveau,
  and Eichinger\/}}]{port01}
Port\'{e}-Agel, F., M.~B. Parlange, C.~Meneveau, and W.~E. Eichinger, 2001:
\newblock A priori field study of the subgrid-scale heat fluxes and dissipation
  in the atmospheric surface layer.
\newblock {\it J. Atmos. Sci.\/}, {\bf 58}, 2673--2698.

\bibitem[{Press et~al.(1992){\it Press, Flannery, Teukolsky, and
  Vetterling\/}}]{pres92}
Press, W.~H., B.~P. Flannery, S.~A. Teukolsky, and W.~T. Vetterling, 1992:
\newblock {\it Numerical Recipes in Fortran\/}.
\newblock Cambridge University Press, 992 pp.

\bibitem[{Sagaut(2001){\it Sagaut\/}}]{saga01}
Sagaut, P., 2001:
\newblock {\it Large Eddy Simulations for Incompressible Flows\/}.
\newblock Springer-Verlag, 426 pp.

\bibitem[{Saiki et~al.(2000){\it Saiki, Moeng, and Sullivan\/}}]{saik00}
Saiki, E.~M., C.-H. Moeng, and P.~P. Sullivan, 2000:
\newblock Large-eddy simulation of the stably stratified planetary boundary
  layer.
\newblock {\it Boundary-Layer Meteorol.\/}, {\bf 95}, 1--30.

\bibitem[{Sarghini et~al.(1999){\it Sarghini, Piomelli, and E.\/}}]{sarg99}
Sarghini, F., U.~Piomelli, and B.~E., 1999:
\newblock Scale-similar models for large-eddy simulations.
\newblock {\it Phys. Fluids\/}, {\bf 11}, 1596--1607.

\bibitem[{Schumann(1991){\it Schumann\/}}]{schu91}
Schumann, U., 1991:
\newblock Subgrid length-scales for large-eddy simulation of stratified
  turbulence.
\newblock {\it Theort. Comput. Fluid Dyn.\/}, {\bf 2}, 279--290.

\bibitem[{Smagorinsky(1963){\it Smagorinsky\/}}]{smag63}
Smagorinsky, J., 1963:
\newblock General circulation experiments with the primitive equations.
\newblock {\it Mon. Wea. Rev.\/}, {\bf 91}, 99--164.

\bibitem[{Sorbjan(1986){\it Sorbjan\/}}]{sorb86}
Sorbjan, Z., 1986:
\newblock Local similarity of spectral and cospectral characteristics in the
  stable-continuous boundary layer.
\newblock {\it Boundary-Layer Meteorol.\/}, {\bf 35}, 257--275.

\bibitem[{Sorbjan(1989){\it Sorbjan\/}}]{sorb89}
Sorbjan, Z., 1989:
\newblock {\it Structure of Atmospheric Boundary Layer\/}.
\newblock Prentice-Hall, 317 pp.

\bibitem[{Stull(1988){\it Stull\/}}]{stul88}
Stull, R.~B., 1988:
\newblock {\it An Introduction to Boundary Layer Meteorology\/}.
\newblock Kluwer Academic Publishers, 670 pp.

\bibitem[{Sullivan et~al.(1994){\it Sullivan, McWilliams, and
  Moeng\/}}]{sull94}
Sullivan, P.~P., J.~C. McWilliams, and C.-H. Moeng, 1994:
\newblock A subgrid-scale model for large-eddy simulation of planetary
  boundary-layer flows.
\newblock {\it Boundary-Layer Meteorol.\/}, {\bf 71}, 247--276.

\bibitem[{Sullivan et~al.(2003){\it Sullivan, Horst, Lenschow, Moeng, and
  Weil\/}}]{sull03}
Sullivan, P.~P., T.~W. Horst, D.~H. Lenschow, C.-H. Moeng, and J.~C. Weil,
  2003:
\newblock Structure of subfilter-scale fluxes in the atmospheric surface layer
  with application to large-eddy simulation modelling.
\newblock {\it J. Fluid Mech.\/}, {\bf 482}, 101--139.

\bibitem[{Townsend(1976){\it Townsend\/}}]{town76}
Townsend, A.~A., 1976:
\newblock {\it The structure of turbulent shear flow\/}.
\newblock Cambridge University Press, 429 pp.

\bibitem[{van~de Wiel(2002){\it van~de Wiel\/}}]{vand02}
van~de Wiel, B., 2002:
\newblock Intermittent turbulence and oscillations in the stable boundary layer
  over land, Ph.D. thesis, Wageningen University, Netherlands.

\bibitem[{Vreman et~al.(1994){\it Vreman, Guerts, and Kuerten\/}}]{vrem94}
Vreman, B., B.~Guerts, and H.~Kuerten, 1994:
\newblock On the formulation of the dynamic mixed subgrid-scale model.
\newblock {\it Phys. Fluids\/}, {\bf 6}, 4057--4059.

\bibitem[{Wong and Lilly(1994){\it Wong and Lilly\/}}]{wong94}
Wong, V.~C., and D.~K. Lilly, 1994:
\newblock A comparison of two dynamic subgrid closure methods for turbulent
  thermal convection.
\newblock {\it Phys. Fluids\/}, {\bf 6}, 1016--1023.

\bibitem[{Yakhot and Orszag(1986){\it Yakhot and Orszag\/}}]{yakh86}
Yakhot, V., and S.~A. Orszag, 1986:
\newblock Renormalization group analysis of turbulence {I}: Basic theory.
\newblock {\it J. Sci. Comput.\/}, {\bf 1}, 3--51.

\bibitem[{Zang et~al.(1993){\it Zang, Street, and Koseff\/}}]{zang93}
Zang, Y., R.~L. Street, and J.~R. Koseff, 1993:
\newblock A dynamic mixed subgrid-scale model and its application to turbulent
  recirculating flows.
\newblock {\it Phys. Fluids. A\/}, {\bf 5}, 3186--3196.

\end{thebibliography}

\end{article}
\end{document}